\documentclass[iop]{emulateapj}
\usepackage{graphicx}
\newcommand{\uv}{\mbox{$u$-$v$}}
\newcommand{\ex}[1]{\mbox{$\times 10^{#1}$}}

\newcommand{\Msol}{\mbox{$M_\sun$}}

\newcommand{\kms}{\mbox{km s$^{-1}$}}
\newcommand{\muas}{\mbox{$\mu$as}}
\newcommand{\muasyr}{\mbox{$\mu$as~yr$^{-1}$}}

\newcommand{\muJb}{\mbox{$\mu$Jy~beam$^{-1}$}}
\newcommand{\Ra}[4]{\mbox{${#1}^{\rm h} \; {#2}^{\rm m} \; {#3}\fs{#4} $}}
\newcommand{\dec}[4]{\mbox{${#1}\arcdeg \; {#2}\arcmin \; {#3}\farcs{#4} $}}

\newcommand{\thfl}{\mbox{$\theta_{\rm90\%\;flux}$}}

\newcommand{\vw}{\mbox{$v_{\rm w}$}}
\newcommand{\Msolxyr}{\mbox{$M_\sun$~yr$^{-1}$}}

\shortauthors{Bietenholz, Bartel \& Rupen}
\shorttitle{SN 1986J VLBI}

\begin{document}
      
\title{SN 1986J VLBI. II. The Evolution of the Shell and the Central Source}

\author{M. F. Bietenholz\altaffilmark{1,2}, N. Bartel\altaffilmark{2}
and M. P. Rupen\altaffilmark{3}}

\altaffiltext{1}{Hartebeesthoek Radio Observatory, PO Box 443, Krugersdorp,
1740, South Africa}
\altaffiltext{2}{Department of Physics and Astronomy, York University, Toronto,
M3J~1P3, Ontario, Canada}
\altaffiltext{3}{National Radio Astronomy Observatory, Socorro, New Mexico
87801, USA}

\slugcomment{Version 8.1 \today}

\begin{abstract}
We present new VLBI images of supernova 1986J, taken at 5, 8.4 and
22~GHz between $t = 22$ to 25 yr after the explosion.  The shell
expands $\propto t^{0.69 \pm 0.03}$\@. We estimate the progenitor's
mass-loss rate at $(4 \sim 10) \times 10^{-5}$~\Msolxyr\ (for $v_{\rm
w} = 10$~\kms).  Two bright spots are seen in the images.  The first,
in the northeast, is now fading.  The second, very near the center of
the projected shell and unique to SN~1986J, is still brightening
relative to the shell, and now dominates the VLBI images.  It is
marginally resolved at 22~GHz (diameter $\sim$0.3~mas;
$\sim5\ex{16}$~cm at 10~Mpc).  The integrated VLA spectrum of SN~1986J
shows an inversion point and a high-frequency turnover, both
progressing downward in frequency and due to the central bright spot.
The optically-thin spectral index of the central bright spot is
indistinguishable from that of the shell.  The small proper motion of
$1500 \pm 1500$~\kms\ of the central bright spot is consistent with
our previous interpretation of it as being associated with the
expected black-hole or neutron-star remnant.  Now, an alternate
scenario seems also plausible, where the central bright spot, like the
northeast one, results when the shock front impacts on a condensation
within the circumstellar medium (CSM)\@.  The condensation would have
to be so dense as to be opaque at cm wavelengths ($\sim 10^3\times$
denser than the average corresponding CSM) and fortuitously close to
the center of the projected shell. We include a movie of the evolution
of SN~1986J at 5~GHz from $t = 0$ to 25~yr.
\end{abstract}


\section{INTRODUCTION}
\label{sintro}

Supernova 1986J was one of the most radio luminous supernovae ever
observed.  Its long-lasting and strong radio emission and its relative
nearness makes it one of the few supernovae for which it is possible
to produce detailed images with very-long-baseline interferometry
(VLBI), in addition to being one of the few supernovae still
detectable more than $t = 20$ years after the explosion.

SN~1986J was first discovered in the radio, some time after the
explosion \citep{vGorkom+1986, Rupen+1987}. The best estimate
of the explosion epoch is $1983.2 \pm 1.1$ \citep[$t = 0$, Bietenholz
et~al. 2002, see also][]{Rupen+1987, Chevalier1987,
WeilerPS1990}\nocite{SN86J-1}.  It occurred in the nearby galaxy
NGC~891, whose distance is $\sim 10$~Mpc \citep[see
e.g.,][]{Tonry+2001, Ferrarese+2000, Tully1988, Kraan-Korteweg1986,
Aaronson+1982}.  We will adopt the round value of 10~Mpc throughout
this paper.

Early optical spectra showed that the supernova was unusual, but based
on its prominent H$\alpha$ lines it was classified as a Type~IIn
supernova \citep{Rupen+1987}.  Later, \citet{Leibundgut+1991}
suggested that it might be of Type Ib.
VLBI observations started soon after discovery \citep{Bartel+1987},
and an image, the first of any optically identified supernova, was
obtained soon after \citep{Bartel+1991}, and showed a heavily
modulated shell structure with protrusions.  VLBI observations at
subsequent epochs up to 1999 led to a series of images showing the
expansion of the shell, and allowing the expansion velocity and
deceleration to be measured \citep{SN86J-1}.

The evolution of the radio spectrum of a supernova is in most cases
predictable.  In particular, once supernovae have become optically thin
after a few months or years, they usually display power-law
radio spectra with spectral indices, $\alpha$, in the range of $-0.8$ to
$-0.5 \; (S \propto \nu^\alpha)$.  SN~1986J also showed such a
spectrum before 1998, but a remarkable change occurred after that.  An
inversion appeared in the spectrum, with the brightness increasing
with increasing frequency above $\sim$10~GHz, and a high-frequency
turnover at $\sim$20~GHz \citep{SN86J-1}.

This spectral inversion was associated with a bright spot in the
(projected) center of the expanding shell, as we showed in
\citet{SN86J-Sci} using phase-referenced multi-frequency VLBI imaging.
At that time (late 2002) the bright spot was clearly present in the
15~GHz image, but not discernible in the 5~GHz one.  We report
here on new VLBI and VLA observations of SN~1986J which show the
further evolution of this unique object.

\section{Observations and Data Reduction}
\label{sobs}

\subsection{VLBI Measurements}
\label{svlbiobs}

\begin{deluxetable*}{l  c c c c }
\tabletypesize{\scriptsize}
\tablecolumns{5}
\tablecaption{VLBI Observations of SN~1986J}
\tablehead{
~~~Date & Frequency & \colhead{Antennas\tablenotemark{a}}
                  & Total time & Recording rate \\
               & (GHz)&     & (hr)  & (Mbits~s$^{-1}$)
}
\startdata
2005 Apr 25 &    22 & VLBA, Ef, Gb, Y27 &   12 & 256  \\
2005 Oct 24 & \phn5 & VLBA, Ef, Gb, Y27, Jb, On, Wb, Tr &   12 & 256  \\
2006 Dec 3 &  \phn8 & VLBA, Ef, Gb, Y27 &   15 & 512  \\
2006 Dec 10 &    22 & VLBA, Ef, Gb, Y27 &   15 & 512  \\
2008 Oct 26 & \phn5 & VLBA, Ef, GB, Y27, Jb, Mc, Nt, Tr, Wb &  18 & 512 \\ 
\enddata
\tablenotetext{a}{
  VLBA, ten 25~m dishes of the NRAO Very Long Baseline Array;\phn
  Ef, 100~m, MPIfR, Effelsberg, Germany;\phn
  Gb, 105~m, NRAO, Green Bank, WV, USA;\phn
  Y27, equivalent diameter 130~m, NRAO, near Socorro, NM, USA;\phn
  Jb,  76~m, Jodrell Bank, UK;\phn
  Mc,  32~m, IdR-CNR, Medicina, Italy;\phn
  Nt,  32~m, IdR-CNR, Noto, Italy;\phn
  On,  20~m, Onsala Space Observatory, Sweden;\phn
  Tr, 32~m, Torun, Poland;\phn
  Wb, equivalent diameter 94~m, Westerbork, the Netherlands.\phn}
\label{tobs}
\end{deluxetable*}

The VLBI observations of SN~1986J were made between 2005 and 2008 with
global VLBI arrays of 13 to 18 antennas and with durations of 12 to 15
hours.  The observations of SN~1986J were interleaved with ones of
3C~66A, only 40\arcmin\ away on the sky, which we used as a
phase-reference source.  Details of the four observing runs are given
in Table~\ref{tobs}. The declination of +42\arcdeg\ of SN~1986J
enabled us to obtain dense and only moderately elliptical
\uv~coverage.  As usual, a hydrogen maser was used as a time and
frequency standard at each telescope. The data were recorded with
either the VLBA or the MKIV VLBI system with sampling rates of 256 or
512~Mbits per second.  We observed at frequencies of 5.0, 8.4 and
22~GHz.  The data were correlated with the NRAO\footnote{The National
Radio Astronomy Observatory, NRAO, is a facility of the National
Science Foundation operated under cooperative agreement by Associated
Universities, Inc.}
VLBA processor in Socorro, NM (USA), and the analysis was carried out
with NRAO's Astronomical Image Processing System (AIPS).  The initial
flux density calibration was done through measurements of the system
temperature at each telescope, and improved through self-calibration
of the 3C~66A data.

We phase-referenced to 3C~66A for all of the present observing
runs. Our positions in this paper are given relative to an assumed
position of RA = \Ra{02}{22}{39}{611500}, decl.\ =
\dec{43}{02}{07}{79884} taken from the International Celestial Reference
Frame, ICRF \citep{Fey+2004}.
for the brightness peak of 3C~66A.

\subsection{VLA Observations}
\label{svla}

In addition to the VLBI observation, we also obtained VLA observations
to measure SN~1986J's total flux density at a range of different
frequencies, both during our VLBI observations and at other times. We
also reduced a number of data sets from the VLA archives in order to
obtain additional flux density measurements.  The epochs and the
frequencies observed are listed in Table~\ref{tflux}.  These
observations allowed us to monitor the evolution of SN~1986J's
integrated radio spectrum.

The VLA data were reduced following standard procedures, with the flux
density scale calibrated by using observations of the standard flux
density calibrators (3C~286 and 3C~48) on the scale of
\citet{Baars+1977}.  An atmospheric opacity correction using mean
zenith opacities was applied, and NGC~891/SN~1986J was self-calibrated
in phase to the extent permitted by the signal-to-noise ratio for each
epoch and frequency.  At lower resolutions the supernova cannot be
reliably separated from the diffuse emission from the galaxy.
Therefore, we cite only flux densities obtained at FWHM
resolutions\footnote{We take the geometric mean of the major and minor
axes of the elliptical Gaussian fit to the dirty beam as our measure
of resolution.}
of $<7$\arcsec.  For the 0.33~GHz measurements we incorporate an
additional uncertainty to reflect the difficulty of accurate galaxy
subtraction.  

We note also that the gradual introduction of EVLA antennas into the
VLA array might cause some problems in the flux-density calibration.
Since the bandpass of the EVLA antennas differs from that of the older
VLA-antennas, non-closing offsets are introduced for continuum
observations such as ours, with the gain of an EVLA-EVLA baseline
being typically several percent higher than that of a VLA-VLA
baseline.  Such non-closing offsets cannot be calibrated out using
only standard procedures, which assume that the gains are dependent
only on the antennas, and not on the baseline.  These closure errors
might compromise the accuracy of the flux density calibration.  To
determine whether this effect is significant, we carried out, for
several epochs and frequencies, a more elaborate flux-density
calibration scheme using baseline-dependent factors calculated by the
AIPS task BLCAL\@.  Although the more elaborate calibration led to an
improvement in the dynamic range, in no case was the normal
flux-density scale in error by more than 1\%.  Since we include in all
cases a flux-density calibration uncertainty of $\ge 5$\%, we can say
that any errors introduced by the use of the EVLA antennas are well
within our stated uncertainties.

\begin{deluxetable*}{l  c c c c c c c }
\tabletypesize{\scriptsize}
\tablecaption{Flux Densities of SN~1986J from VLA Observations}
\tablehead{
\colhead{Date} & \multicolumn{6}{c}{Flux densities\tablenotemark{a}}\\
               & \colhead{0.32~GHz} & \colhead{1.43 GHz}& \colhead{4.9 GHz\tablenotemark{b}}
               & \colhead{8.4~GHz\tablenotemark{c}}
               & \colhead{14.9 GHz} & \colhead{22 GHz\tablenotemark{d}}
               & \colhead{43.3~GHz} 
}
\startdata
1995 Jun 22\tablenotemark{e} 
             &           &$25.6 \pm1.3 $&$14.2 \pm 0.8$&            &$9.1 \pm 0.7$&$8.0 \pm 0.8$\\
1996 Oct 25\tablenotemark{e} &  &       &$11.0 \pm 0.6$&$8.7 \pm0.5$\\
1997 Jan 31\tablenotemark{e} &  &       &              &$8.2 \pm0.5$\\
1998 Feb  9\tablenotemark{e} &  &       &              &            &$5.65\pm0.69$&$9.3 \pm 1.5$\\
1998 Jun  5             &  &$13.7\pm0.9$\tablenotemark{f}
                                        &$ 8.5 \pm 0.5$&$7.2 \pm 0.4$\\
1999 Feb  2             &  &            &$ 7.3 \pm 1.5$&$6.1 \pm 0.4$&$6.1 \pm 0.1$&$9.5 \pm 1.0$\\
1999 Jun 13\tablenotemark{e} & 
                        &$11.69\pm0.62$&             &$5.68\pm0.30$&$4.92\pm0.52$& \\
2000 Oct 19\tablenotemark{e} &   
                        & $8.97\pm0.47$&$5.43\pm0.29$&$4.76\pm0.26$&$5.88\pm0.42$&$5.67\pm0.65$ \\
2001 Jan 25\tablenotemark{e} &     
                        & $9.02\pm0.22$&             &$4.75\pm0.36$\\
2002 Jan 27\tablenotemark{e}
            &           & $7.68\pm0.46$&$4.15\pm0.26$&$4.13\pm0.28$&$5.02\pm0.54$&$4.22\pm0.44$ \\
2002 May 25 &           & $7.10\pm0.70$&$4.20\pm0.30$&$3.80\pm0.30$&$4.70\pm0.50$
                                                     &$5.00\pm0.60$&$3.9^{+2.1}_{-0.9}$ \\
2002 Nov 11 &           &              &$3.86\pm0.28$&$4.16\pm0.25$&             &$5.37\pm0.50$\\
2003 Jan 8  &           &              &$3.75\pm0.45$&$3.75\pm0.30$&$4.25\pm0.44$&$4.41\pm0.48$&$3.90\pm0.50$\\
2003 Jun 22 &           & $6.63\pm0.57$&$3.50\pm0.25$&$3.88\pm0.20$&$4.57\pm0.70$& \\
2004 Jan 31\tablenotemark{e}
            &           &              &             &             &$4.72\pm0.29$&$4.74\pm0.27$\\
2004 Sep 23 &           & $4.55\pm0.33$&$3.07\pm0.20$&$3.45\pm0.20$&$4.44\pm0.26$&$4.79\pm0.49$&$3.59\pm0.40$\\
2005 Jan  8 &           & $4.49\pm0.33$&$2.91\pm0.16$&$3.21\pm0.19$&$4.01\pm0.27$&$4.11\pm0.23$&$3.18\pm0.38$\\
2005 Apr 26 &           &              &$2.58\pm0.43$&$2.92\pm0.17$&$3.14\pm1.12$&$3.73\pm0.28$\\
2005 Jun 13\tablenotemark{e}
            &           &              &$2.45\pm0.17$&$3.11\pm0.19$&$4.01\pm0.34$&$3.75\pm0.59$\\
2005 Oct 10 &           &              &             &             &             &$4.25\pm0.55$\\
2006 Mar 6  &           & $3.99\pm0.26$&$2.62\pm0.17$&$3.20\pm0.16$&$3.92\pm0.26$&$3.93\pm0.20$&$2.91\pm0.35$\\
2006 Jun 6  &           &              &$2.45\pm0.15$&$3.20\pm0.17$&$4.16\pm0.35$&$3.61\pm0.21$&$2.32\pm0.34$\\
2006 Aug 30\tablenotemark{e}     &  &  &$2.12\pm0.13$&             &             &$3.24\pm0.43$\\
2006 Dec 4  &           &              &             &$3.13\pm0.31$&\\
2006 Dec 11 &           &              &             &             &             &$4.54\pm0.91$\\
2007 Apr 23 &           &              &             &             &$3.69\pm0.29$&$3.57\pm0.20$&$2.07\pm0.29$\\
2007 Jul 31 &           &              &$2.06\pm0.15$&             &$3.77\pm0.32$\\
2007 Aug 19\tablenotemark{g}
            &$8.9\pm1.8$& $3.07\pm0.13$&             &$2.97\pm0.12$&             \\
2007 Aug 23\tablenotemark{h} & &       &             &             &             &$3.12\pm0.19$&$2.13\pm0.23$ \\
2008 Oct 26 &           &              &$2.27\pm0.12$&  \\
\enddata
\tablenotetext{a}{Flux densities from fitting images and/or fitting
\uv~plane models.  The tabulated uncertainties include a systematic
uncertainty in the flux density calibration of 5\% or more at
frequencies of 22~GHz and below, and 10\% or more at 43~GHz.
They also include the estimated uncertainty of the galaxy
subtraction for those epochs/frequencies at which the galaxy
contribution was significant. We only tabulate measurements
at resolutions $<7$\arcsec, and ones where the galaxy contribution
was $<50$\% of the brightness of SN~1986J.}
\tablenotetext{b}{Observations between 4.86 and 4.99 GHz.}
\tablenotetext{c}{Observations between 8.41 and 8.46 GHz.}
\tablenotetext{d}{Observations between 22.21 and 22.66 GHz.}
\tablenotetext{e}{Data from the VLA archives.}
\tablenotetext{f}{Flux density at 1.7 GHz, rather than 1.43 GHz.}
\tablenotetext{g}{Values are the average from two observing runs on 2007 Jul 31 and 2007 Sep 6.}
\tablenotetext{h}{Values are the average from two observing runs on 2007 Jul 31 and 2007 Sep 15.}
\label{tflux}
\end{deluxetable*}

In Table~\ref{tflux} we list the total flux density measurements that
we obtained from VLA observations.  Note that some of these values
have been previously published \citep[see][]{SN86J-COSPAR-2, SN86J-COSPAR,
SN86J-Sci}, but we repeat them here for the convenience of the reader
and to allow a better interpretation of the evolution of SN~1986J's radio
emission.

\begin{figure*}
\centering
\includegraphics[height=0.67\textheight]{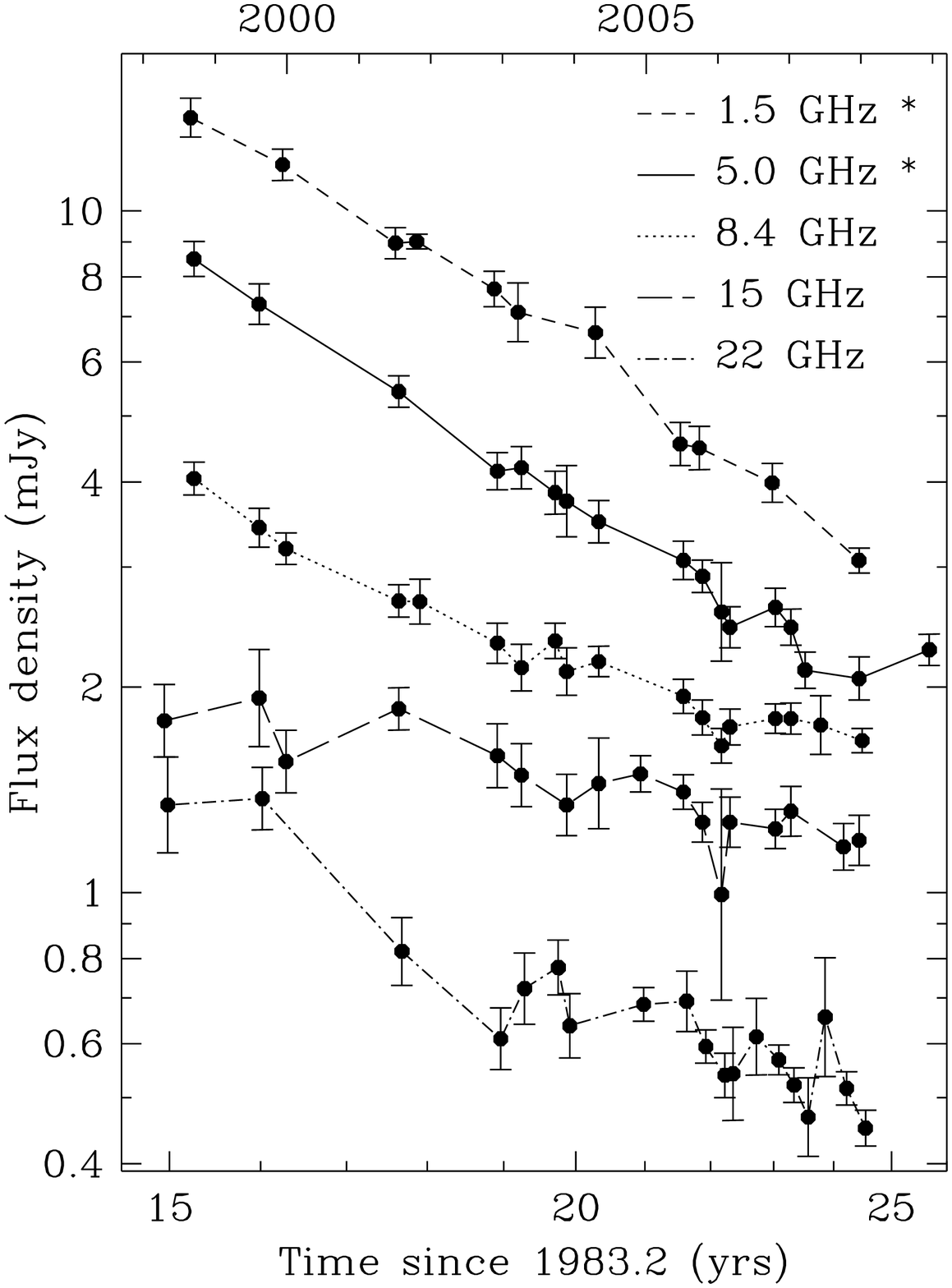} 
\caption{Multi-frequency radio lightcurves for SN~1986J, as determined
from VLA observations. We plot the lightcurves at 1.4, 5.0, 8.4, 15,
and 22~GHz.  The flux-density scale on the left axis is relevant
directly for the lightcurves at 1.5 and 5.0~GHz (marked with an asterisk),
while those at 8.4, 15 and 22 GHz are shifted logarithmically
downwards for better visibility, so that the shape of the lightcurve
is preserved even though the flux density scale is not.  The curves
are shifted by the following factors: 8.4~GHz by 0.56, 15~GHz by 0.32,
and 22~GHz by 0.14.  The uncertainties are estimated standard
errors, including statistical and systematic contributions.  The data
include both our own and re-reduced archival data (see
Table~\ref{tflux}).}
\label{flightcurve}
\end{figure*}

\begin{figure*}
\centering
\includegraphics[height=0.67\textheight]{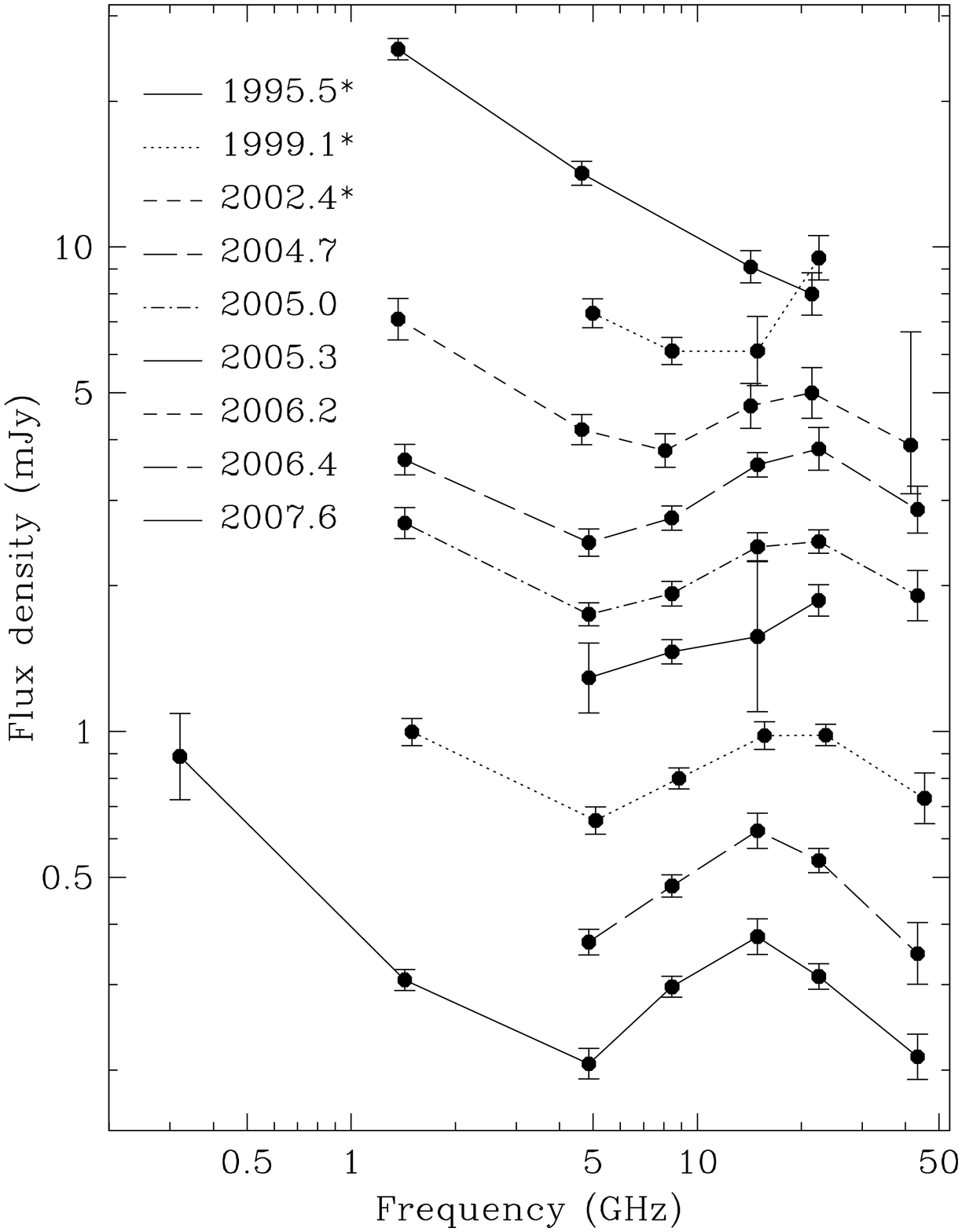} 
\caption{The evolving radio spectrum of SN~1986J, as determined from
VLA observations.  Each curve shows the radio spectrum at the epoch
indicated at left, with the earliest spectrum at the top.  The
flux-density scale on the left axis is relevant directly for the first
three curves from the top (1995.5, 1999.1 and 2002.4), whose epochs
are marked with an asterisk.  The other curves are shifted
logarithmically progressively downwards for better visibility, so that
the spectral shape is preserved even though the flux density scale is
not.  The curves are shifted by the following factors: 2004.7 by 0.80,
2005.0 by 0.60, 2005.3 by 0.50, 2006.2 by 0.25, 2006.4
by 0.15, and 2007.6 by 0.10.  The uncertainties are estimated
standard errors, including statistical and systematic contributions.
The data include both our own and re-reduced archival data (see
Table~\ref{tflux}).}
\label{fspectra}
\end{figure*}

We plot a selection of these flux densities in two ways: first, in
Figure~\ref{flightcurve}, we plot the evolution of the flux-density as
a function of time at several different frequencies. Second, in
Figure~\ref{fspectra} we plot the radio spectrum as a function of time
at a number of epochs.  We focus in this paper on the flux-density
evolution in the last decade, for lightcurves covering the earlier
part of SN~1986J's evolution, see \citet{SN86J-1} and
\citet{WeilerPS1990}.

Figure \ref{flightcurve} shows that the flux density at all our
observing frequencies is generally decreasing.  In terms of a
power-law decay with $S_\nu \propto t^\beta$,
a weighted least-squares fit gives an average value of $\beta = -2.2$
over the last decade and over the frequencies of 1.5, 5.0, 8.4, 15 and
22~GHz, with the fastest decay seen at 1.5~GHz with $\beta = -3.3$
and the slowest at 15~GHz with $\beta = -1.0$.

\pagebreak[4]

~

\subsection{Radio Spectral Index}
\label{sspix} 

As is clear from Figure~\ref{fspectra}, the radio spectral index,
$\alpha$ is a strong function of both time and frequency.  Up to
1995.5, however, the spectrum at least between 1.4 and 15~GHz seems
well described by a single power-law.  Earlier on, up to 1988,
\citet{WeilerPS1990} reported a spectral index for the optically-thin
part of $-0.67_{-0.04}^{+0.08}$.  \citet{BallK1995}, for the period up
to 1991, find a somewhat flatter value of $-0.57 \pm 0.01$ from
fitting a diffusive acceleration model.
Fitting a power-law to the first set of flux densities included in
Table~\ref{tobs}, namely those of 1995.5, we find a notably flatter
spectrum, with a spectral index of $-0.44 \pm 0.03$ (we note that if
we drop the 22~GHz flux density from this fit on the suspicion that it
might already be contaminated by the inversion, the fitted spectral
index changes by less than the uncertainties).  So the spectrum seems
to have flattened significantly between 1988 and 1995.

After 1995, the inversion appears in the spectrum, and its evolution
becomes more complex.  We will focus first on the low-frequency part
of the spectrum, which we associate with the shell.  In
\citet{SN86J-1}, we performed a two-part decomposition of the spectrum
for the period 1998 to 2002, and obtained a spectral index of
$-0.55_{-0.16}^{+0.09}$.
For even more recent times, we can calculate some representative
spectral indices from the flux densities in Table~\ref{tobs}.  
From our measurements on 2007 Aug 19, we find $\alpha_{\rm 0.3
GHz}^{\rm 1.4 GHz} = -0.71 \pm 0.14$.  Although the uncertainties are
rather large, this suggests also that the part of the spectrum at the
lowest frequencies is again becoming steeper since 1995.5.

In summary, the evidence suggests that the part of the spectrum below
the inversion, which we associate with the shell, was fairly steep
with $\alpha \simeq -0.62$ at times till 1988, flattening to $\alpha
\simeq -0.44$ in 1995, and once again steepening somewhat to $\alpha
\simeq -0.7$ since then.  By averaging the spectral indices derived by
us and others for the low-frequency/optically-thin part of the
spectrum, we find an average value over the whole timerange of $-0.61$
with an rms of 0.11.

Another region of interest is the spectrum {\em above} the
high-frequency turnover.  This part of the spectrum is dominated by
emission from the central bright spot, and thus presumably represents
its unabsorbed spectrum. A weighted average between 2006 Mar 6 and
2007 Aug 23 gives $\alpha_{\rm 22 GHz}^{\rm 44 GHz} = -0.61 \pm 0.11$.

\section{VLBI Images}
\label{simages}

\begin{figure*}
\centering
\includegraphics[height=0.8\textheight]{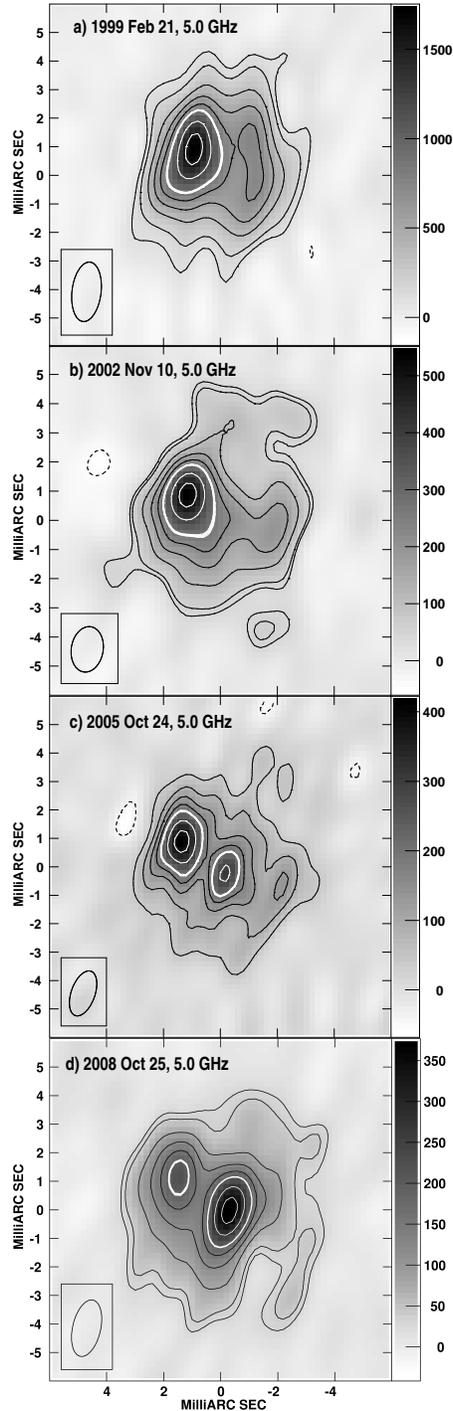}
\caption{\small A sequence of VLBI images of SN~1986J at 5~GHz,
showing its evolution over the last decade.  In each panel, the
coordinate system is centered on the estimated position of the center
of the shell (\Ra{02}{22}{31}{321457}, \dec{42}{19}{57}{25951}, see
text).  The contours are in \% of the peak brightness, with negative
ones being dotted.  Each panel has contours at 10, 20, 30, 40, {\bf
50}, 70, and 90\%, with the 50\% one being emphasized and the lowest
white one.  The lowest contours, given for each panel below, are three
times the rms background brightness.  The FWHM size of the convolving beam
is indicated at lower left in each panel.  North is up and east is to
the left. This figure is also available as an mpeg animation in the
electronic edition of the {\em Astrophysical Journal}.
a) 1999 Feb 21, 5 GHz.  The peak brightness was 1720~\muJb, the
lowest contour 5.5\%, the rms background brightness 31~\muJb,
and the convolving beam $2.09 \times 1.02$~mas at p.a.\ $-7$\arcdeg.
b) 2002 Nov 10, 5 GHz.  The peak brightness was 543~\muJb, the
lowest contour 7.6\%, the rms background brightness 14~\muJb,
and the convolving beam  $1.57 \times 1.12$~mas at p.a.\ $-10$\arcdeg.
The visibility data were tapered slightly in $u$ to increase
signal-to-noise and produce a somewhat less elongated beam. 
c) 2005 Oct.\ 24, 5.0 GHz.  The peak brightness was 414~\muJb, the
lowest contour 9.3\%, the rms background brightness 13~\muJb,
and the convolving beam $1.65 \times 0.83$~mas at p.a.\
$-20$\arcdeg.
d) 2008 Oct.\ 25, 5.0 GHz.  The peak brightness was 388~\muJb, the
lowest contour 7\%, the rms background brightness 9.0~\muJb, and the
convolving beam was $2.02 \times 0.98$~mas at p.a.\ $-13$\arcdeg.}
\label{fimage5G}
\end{figure*}

\begin{figure*}
\centering
\includegraphics[width=0.473\textwidth]{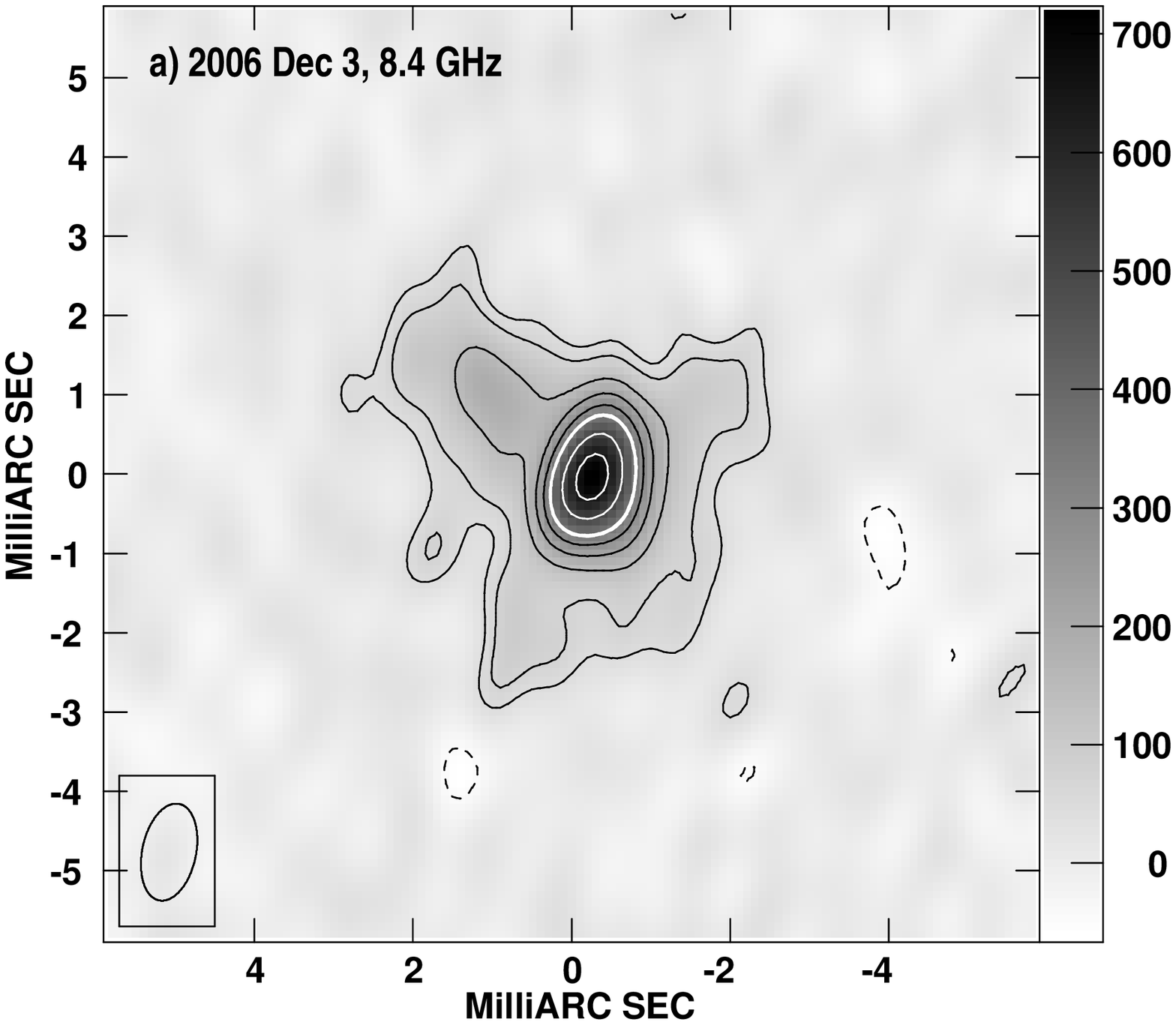}
\includegraphics[width=0.49\textwidth]{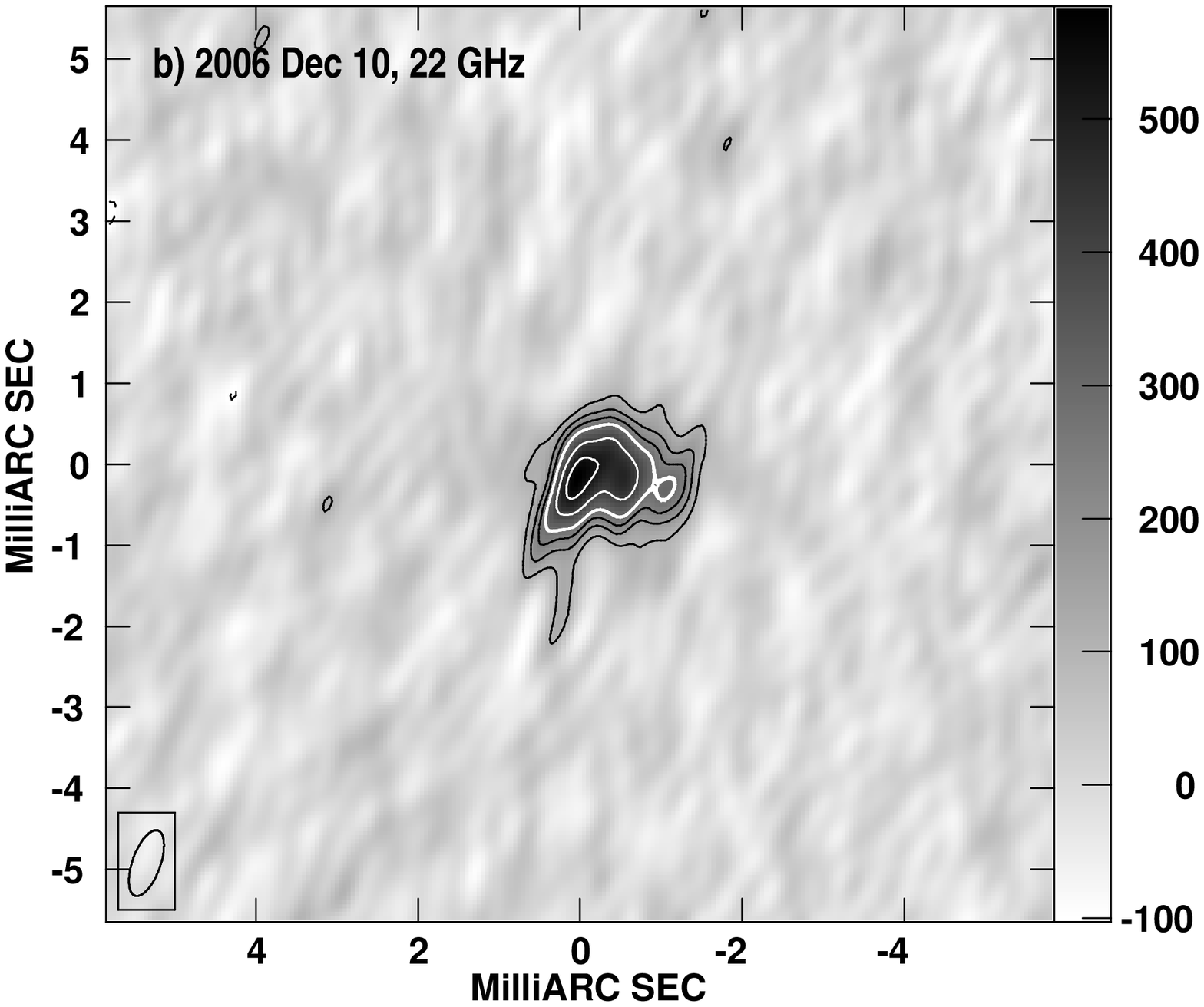}
\caption{VLBI images of SN~1986J at 8.4 and 22~GHz.
In each panel, the coordinate system is centered on the estimated
position of the center of the shell (\Ra{02}{22}{31}{321457},
\dec{42}{19}{57}{25951}, see text).  The contours are in \% of the
peak brightness, with the lowest being at three times the rms
background brightness, and the 50\% being emphasized and being the
first white contour.  The FWHM of the convolving beam is shown at
lower left.  North is up and east to the left.
a) Contours are drawn at $-6$, 6, 10, 20, 30, {\bf 50}, 70, and
90\% of the peak brightness, which was 717~\muJb.  The rms background
brightness was 15~\muJb, and the convolving beam FWHM was $1.24 \times
0.68$~mas at p.a.\ $-12$\arcdeg.
b) Contours are drawn at $-17$, 17, 30, {\bf 50}, 70, and 90\% of
the peak brightness, which was 580~\muJb.  The rms background
brightness was 34~\muJb, and the convolving beam FWHM was $0.85 \times
0.35$~mas at p.a.\ $-20$\arcdeg.}
\label{fimage22}
\end{figure*}

In Figures~\ref{fimage5G} and \ref{fimage22} we show the VLBI images of
SN~1986J at the different frequencies and dates in Table~\ref{tobs}.
For comparison, we include in Figure~\ref{fimage5G} also some of our
previously published images from 1999 and 2002.
We discuss the images in turn, first those at 5~GHz, which
together with the earlier images present the best picture of the
evolution of the supernova, and then those at 8.4 and 22~GHz.

Since all the observations in this paper were phase-referenced to the
brightness peak of 3C~66A, which is expected to be closely related to
the core of the galaxy, we can accurately align the images of SN~1986J
at different times and frequencies to facilitate inter-comparison.
Unfortunately, since the earliest epochs of VLBI observations of
SN~1986J \citep[between 1987 and 1992, see][]{SN86J-1} were not
phase-referenced, we cannot pinpoint the explosion location
accurately, as we did for SN~1993J \citep[see][]{SN93J-1}.  However,
we can visually estimate the position of the center of the shell
sufficiently accurately for our purposes.  We use this center position
as the origin of our coordinate system in all the images, and also as
our estimate of the explosion location.

In particular, we estimated by eye the coordinates of the center of
the shell in each of the four phase-referenced, 5-GHz images shown in
Figure~\ref{fimage5G}. We then averaged these four sets of coordinates
and use the average to represent the coordinates of the shell center
and the explosion center position, which is Ra =
\Ra{02}{22}{31}{321457}, decl.\ = \dec{42}{19}{57}{25951}.  We note
that a minimum uncertainty in this center position can be derived from
the scatter of the individual estimates, and is $\sim 60 \; \mu$as and
$\sim 180 \; \mu$as in RA and decl.\ respectively, and we estimate
realistic uncertainties in the actual position of the center of the
shell the order of 200~\muas.  We note that the largest component of
this uncertainty is not astrometric, but comes from the difficulty of
identifying the center position of the shell in the morphology, and
our measurements of proper motion below do not depend on this estimate
of the position of the shell's center.

\noindent We discuss the images in turn:  
\begin{trivlist}

\item{a) 5 GHz, 1999 Feb.\ 21:} (Fig.~\ref{fimage5G}a) The northeast
bright spot dominates.  There is a suggestion that it has an extension
towards the center.  A version of this image was previously published
in \citet{SN86J-1}.

\item{b) 5 GHz, 2002 Nov.\ 10:} (Fig.~\ref{fimage5G}b) The northeast
bright spot continues to dominate the image.  Again, a slight
extension towards the center is seen, but it cannot be clearly
identified as the central bright spot.  A version of this image was
previously published in \citet{SN86J-Sci}.

\item{c) 5 GHz, 2005 Oct.\ 24:} (Fig.~\ref{fimage5G}c) The central
bright spot is now clearly seen at 5~GHz although it is not yet
as strong as the northeast bright spot. A version of this image was
previously published in \citet{SN86J-COSPAR-2}.

\item{d) 5 GHz, 2008 Oct.\ 25:} (Fig.~\ref{fimage5G}d) The central
bright spot now dominates at 5~GHz, and the northeast bright spot
is becoming relatively fainter.  The shell continues to expand (see
\S~\ref{sexp} below for a determination of the expansion velocity).

\item{e) 8.4 GHz, 2006 Dec.\ 3:} (Fig.~\ref{fimage22}a) This run
benefited from the newly available 512~Mbits~s$^{-1}$ recording
bandwidth.  The resulting image is dominated by the central bright
spot, with the northeast bright spot and the remainder of the shell
being only marginally discernible.

\item{f) 22 GHz, 2005 Apr.\ 25:} This image was very similar to the
later 22~GHz image in 2006, discussed below.  As it was of notably
lower signal-to-noise ratio, we do not reproduce it here \citep[it was
reproduced in][]{SN86J-COSPAR-2}.

\item{g) 22 GHz, 2006 Dec.\ 10:} (Fig.~\ref{fimage22}b) At this
frequency, the steep-spectrum shell is not visible, and the central
bright spot dominates.  The latter is now clearly resolved and appears
somewhat asymmetrical, with a brightness peak to the east.  The
finger-like extension to the south is of marginal significance.  This
image was deconvolved with CLEAN, but we note that maximum entropy
deconvolution (AIPS task VTESS) produces a very similar image.
\end{trivlist}

\subsection{Expansion}
\label{sexp}

The average expansion velocity and deceleration can be determined from
the VLBI data by estimating the size of the supernova at each epoch.
As SN~1986J's morphology is becoming increasingly irregular,
estimating the size by fitting a simple geometrical model to the
visibility data \citep[as we did for SN~1993J, see e.g.,][]{SN93J-1,
SN93J-2} is no longer practical.  We take as a representative
(angular) radius the average radius of the contour which contains 90\%
of the total flux density in the image.  Measures of the radius
derived from images are complicated by the convolution of the images
by a clean beam, but a consistent estimate of the expansion rate can
be obtained by comparing radii derived from images convolved with an
approximately co-evolving beam.  For a fuller discussion, see
\citet{SN86J-1}, where we showed that the expansion up to 1999 had a
power-law form with the radius, $r \propto t^{0.71 \pm 0.11}$.  We
accordingly use images convolved with a co-moving beam,\footnote{This
is the same
procedure we used in \citet{SN86J-1}, where we also found that the
expansion curve is not sensitive to the exact time-dependence of the
restoring beam, as long as it is similar to the time-dependence of the
radius. The use of an approximate value for the coefficient of the
expansion power-law should therefore not significantly bias our
result.  Note that the value of 0.70 we use for this coefficient is
slightly different from the value of 0.75 that we used in
\citet{SN86J-1}.  The present value is a somewhat better match to the
fit power-law expansion of the supernova.  Note also that our present
value of the expansion coefficient is in agreement within the
uncertainties with the value we had obtained in \citet{SN86J-1}, but
does not support the higher value of $0.90 \pm 0.06$ reported by
\citet{Perez-Torres+2002}. }
whose dimensions scale with $t^{0.70}$.
We call this radius \thfl.  Table~\ref{tthfl} gives the derived radii
for our various observing epochs.  For observations after 2000, we do
not determine radii from the observations at 8~GHz or above because at
these frequencies, the shell was not sufficiently distinct for a
useful measurement.  We further note that the increasingly
non-self-similar evolution of the morphology makes such average
measures of the radius and expansion less reliable, however, our fit
to the expansion should give a reasonable estimate of the properties
of the expansion averaged over the entire shell.

We fix the explosion date, $t_0$, at 1983.2, the value that we
determined in \citet{SN86J-1} from both the expansion measurements and
the radio light-curve.  A least-squares fit of a power-law then gives
an average expansion $\thfl \propto t^{0.69 \pm 0.03}$, consistent
with, but more accurate than the value we had derived in
\citet{SN86J-1} for the measurements up to 1999.

\begin{deluxetable*}{l  c c c c }
\tabletypesize{\small}
\tablecaption{Angular Sizes of SN~1986J}
\tablewidth{0.7\textwidth}
\tablehead{
\colhead{Date} & \colhead{Frequency} 
               & \colhead {Age\tablenotemark{a}}
               &\colhead{Angular Size\tablenotemark{b}}\\
               & \colhead{(GHZ)} 
               & \colhead{(yrs)}
               & \colhead{(mas)} 
}
\startdata
1988 Sep 29 & 8.4 & \phn 5.55 & $1.37 \pm 0.10$\tablenotemark{c} \\
1990 Jul 21 & 8.4 & \phn 7.35 & $1.83 \pm 0.10$\tablenotemark{c} \\
1999 Feb 22 & 5.0 & 15.94     & $3.07 \pm 0.10$\tablenotemark{c} \\
2002 Nov 10 & 5.0 & 19.66     & $3.60 \pm 0.14$ \\
2005 Oct 24 & 5.0 & 22.62     & $3.71 \pm 0.15$ \\
2008 Oct 26 & 5.0 & 25.62     & $4.23 \pm 0.17$ \\
\enddata
\tablenotetext{a}{The explosion date is not precisely known.  We use a
best-fit explosion epoch of 1983.2 determined in \citet{SN86J-1}.}
\tablenotetext{b}{We take as a representative angular radius \thfl,
which is the average radius of the area containing 90\% of the clean
flux density in the image.  The images are convolved with an
approximately co-moving restoring beam.  See text for details.}
\tablenotetext{c}{These values are from data discussed in
\citet{SN86J-1}, although the present values of \thfl\ differ slightly
from the ones reported there because we are using a slightly different
co-moving convolving beam.  In that earlier reference, we used a
convolving beam whose size increased $\propto t^{0.75}$, whereas the
present values are based on a beam whose size increases $\propto
t^{0.70}$, which better matches the actual expansion of the
supernovae.  The difference in derived radii are small, with the
values of \thfl\ changing by less than their stated uncertainties.}
\label{tthfl}
\end{deluxetable*}

\begin{figure*}
\centering
\includegraphics[height=4.in]{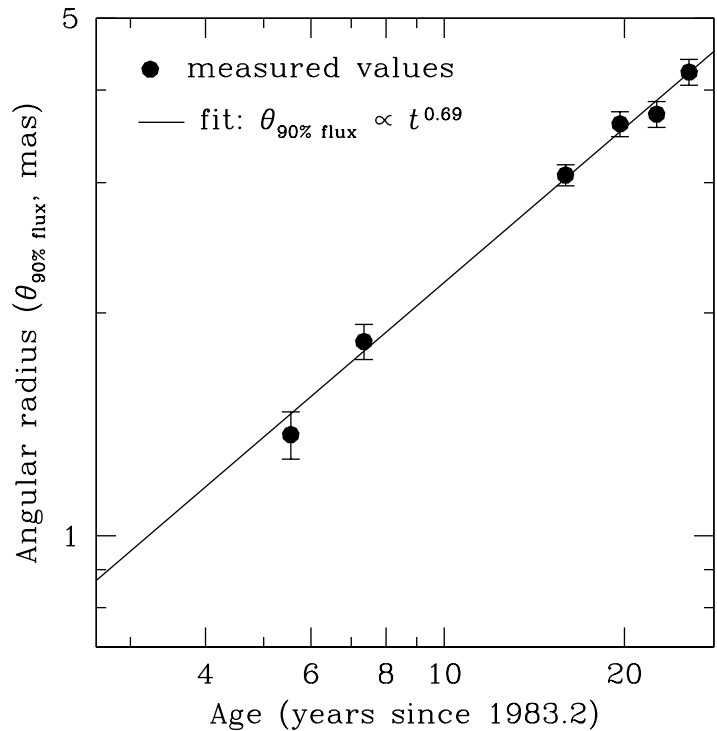}
\caption{The expansion curve of SN~1986J\@.  We plot the angular outer
radius, \thfl\ (the angular radius which contains 90\% of the total
flux density in the image, see text, \S~\ref{sexp} for details),
against the age, both on logarithmic scales.  The line indicates
expansion of the form $\theta \propto (t - t_0)^m$, with $m = 0.69$,
where $t = 0$ corresponds to the fixed epoch 1983.2, which was the
best weighted least-squares fit to the points plotted.}
\label{fexp}
\end{figure*}

We plot the resulting expansion curve in Figure~\ref{fexp}. The two
radii after epoch 1999 fall slightly below the earlier power-law fit,
suggesting a possible slight increase in the deceleration.  However,
we caution that the increase in brightness of the central bright spot
relative to that of the shell might lead to a \thfl\ being a slight
underestimate of the true radius of the shell.

The average expansion velocity measured between the 1999 and 2008
epochs was $5700 \pm 1000$~\kms.
Although this velocity, based on \thfl, is likely not the exact
velocity of the outer radius of the shell of radio emission, it should
be a reasonable approximation thereof.

\subsection{Positions and Proper Motion of the Two Bright Spots}
\label{svspot}

How does the expansion of the supernova compare with the proper
motions of the northeast and central bright spots?  Since our images
are phase-referenced to the brightness peak of the un-related nearby
compact radio source, 3C~66A, we can accurately determine the proper
motion of the two bright spots relative to 3C~66A.

Before we proceed to the proper motions, we briefly discuss the
uncertainties which will apply to such measurements \citep[for other
discussions on the astrometric accuracy obtainable with VLBI,
see][]{SN93J-1, M81-2004, PradelCL2006}. Our estimated astrometric
uncertainties below consist of the following three components: 1) An
uncertainty due to the noise in the image.  2) An uncertainty in
identifying a particular point in the morphology of the reference source,
3C~66A, which is slightly time and frequency
dependent\footnote{\citet{Kovalev+2008a} show that
frequency-dependent shifts in core position are common among compact
radio sources.  For example, see \citet{M81-2004} and \citet{M81-2000}
for measurements of the variability in time and frequency of the
compact core of the nearby galaxy M81.}.
We estimate the latter uncertainty component as 20~\muas\ in RA and
70~\muas\ in decl., with the decl.\ component being larger because the
source is elongated approximately in the north-south direction.  3) An
astrometric uncertainty due to errors in the station coordinates,
earth orientation, and tropospheric correction, which we estimate at
30~\muas\ in each coordinate \citep[see][]{PradelCL2006}.

We first turn to the northeast bright spot.  It was clearly detected
already in 1999 \cite[Bietenholz et al., 2002; for a similar image
from an independent reduction of our observations,
see][]{Perez-Torres+2002}\nocite{SN86J-1}.  We find that, based on a
linear fit to the position in the 1999 Feb., 2002 Nov., 2005 Oct., and
2008 Oct.\ images at 5~GHz, the proper motion of the northeast bright
spot was $60 \pm 20$~\muasyr, corresponding to a speed of $3000 \pm
1000$~\kms, 
$2400 \pm 250$ and $1500 \pm 700$~\kms,
in RA and decl.\ respectively.

\begin{figure*}
\centering
\includegraphics[angle=-90, width=0.58\textwidth]{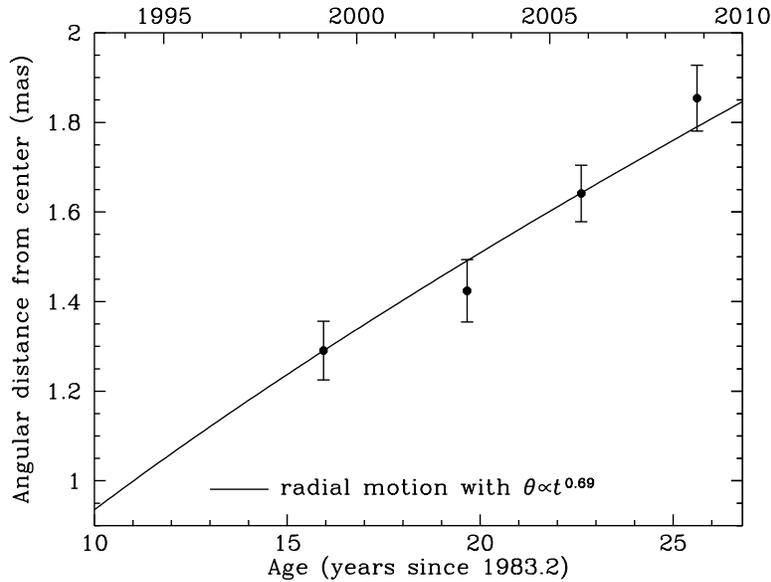}
\caption{The angular distance of the northeast bright spot from the
center as a function of time.  The additional axis labels along the
top give the calendar year. The spot's angular distance is taken to be
that between the spot local maximum and the estimated explosion center
(see text, \S~\ref{svspot}, for details on the estimation of the
explosion center).
The shell expands with outer angular radius $\propto t^{0.69}$ (see
text \S~\ref{sexp}).  Of the possible expansion curves which have
$\theta \propto t^{0.69}$, we show as the solid line the one that best
matches the spot's motion.  Note that the spot's displacement as a
function of time is somewhat dependent upon the choice of the
explosion center, although for reasonable choices, the nature of the
expansion curve remains unchanged. Our plotted error bars are the
uncertainty in the spot position and an estimated uncertainty of
0.1~mas in the explosion location added in quadrature.  The spot's
motion is consistent with homologous expansion.}
\label{fvspot}
\end{figure*}

Does the current proper motion of the northeast bright spot place it
at the explosion center at the time of explosion?  To answer this
question requires knowing the position of the explosion enter.  As
mentioned earlier, this center is not accurately known for SN~1986J,
but as an estimate we use the average position of the center of the
shell, as described in \S~\ref{simages} above.  Extrapolating the
northeast bright spot's present proper motion the explosion epoch of
1983.2 yields a position consistent with the estimated center position
within 1.9 and 0.8 sigma in RA and decl., respectively.  If allowance
for deceleration is made, then the spot's 1983.2 extrapolated position
is even closer to the center.  In fact, a radial motion with the same
power-law dependence on time as the overall expansion, which has $r
\propto t^{0.69}$, is compatible with our measurements, as can be seen
in Figure~\ref{fvspot}, where we show the radial motion of the
northeast bright spot.

A homologous, power-law expansion which includes the northeast bright
spot is therefore consistent with our measurements, with the spot's
present position and proper motion being compatible with it being at
the center of the shell at the explosion epoch.  We note, however,
that the uncertainties are not small, so considerable deviation from
the power-law expansion, for example stronger deceleration, or motion
with a constant speed, are also consistent with our measurements.

We turn now to the central bright spot.  It is best defined in the
images from 2003 June 22 \citep[15~GHz,][]{SN86J-Sci}, 2006 Dec.\ 10
(22~GHz) and 2008 Oct.\ 26 (5~GHz).  If we estimate the center bright
spot's position at each epoch by the position of the brightness peak,
we can determine the spot's proper motion.  A linear fit to the three
positions gives a proper motion of +4 and +31~\muasyr\ in RA and
decl., respectively.  
The morphology of the bright spot makes the peak brightness position
somewhat resolution-dependent, (and possibly frequency-dependent), and
we estimate that the accuracy with which any particular spot in the
morphology can reliably be identified to be on the order of 100~\muas.
In addition to the other uncertainties estimated above, we use a
rounded value of 120~\muas\ for the uncertainty in the spot position
in each coordinate, which leads to an uncertainty in the proper motion
of 31~\muasyr\ in each coordinate.  This implies a speed of $1500 \pm
1500$~\kms\ at 10~Mpc. The estimate of the proper motion is consistent
with the central bright spot being stationary.  The average position
of the central bright spot is $300 \pm 150$~\muas\ from the geometric
center of the shell (see \S~\ref{simages}).
However, given the uncertainty in associating the explosion center
with the geometric center of the shell, we only say that the position
of the central bright spot is marginally consistent with that of the
explosion center.

\begin{figure*}
\centering
\includegraphics[width=0.58\textwidth]{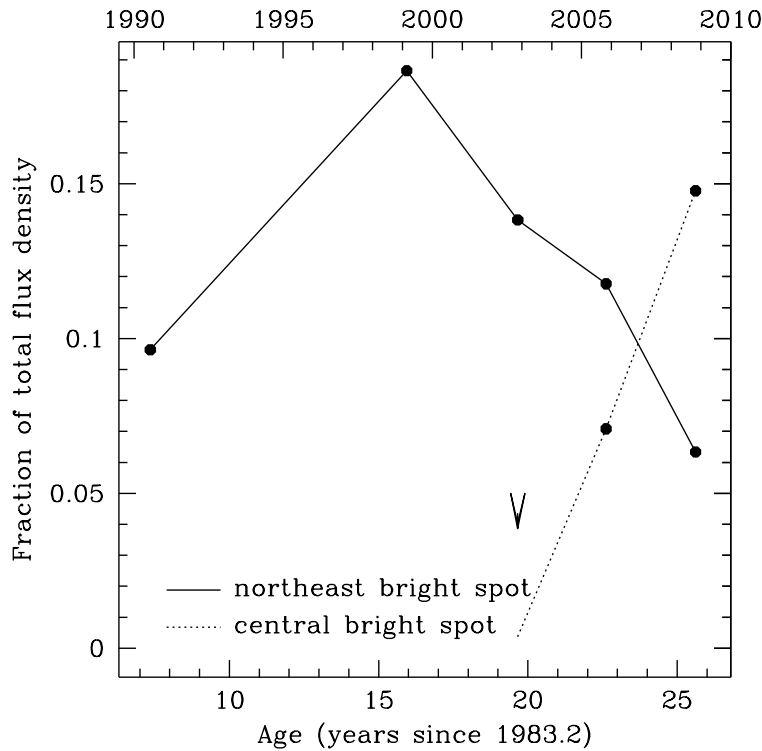}
\caption{The evolution at 5~GHz of the fractional flux density of the
bright spots in the northeast and in the center of the projected shell
as a function of time.  The additional axis labels along the top give
the calendar year.  The bright spot flux densities are estimated by
fitting, by least squares, a point source and a baseline level in the
region near the spot.  The fraction of the flux density is taken to be
the fitted flux density for the point source divided by the total
CLEAN flux density in the image.  For the first three epochs, the
bright spot in the center is not discernible.  The chevron shows an
upper limit on the central bright spot flux density at age 19.7~yr,
which we derived by taking half the surface brightness at its
location.}
\label{ffspot}
\end{figure*}

\subsection{Brightness Evolution of the Two Bright Spots}
\label{sfspot}

In order to quantify the evolution of the brightness of the northeast
and central bright spots, we fitted simple models to each bright spot
in the image plane.  In particular, in the region near each spot we
fit an elliptical Gaussian of the same dimensions as the restoring
beam, i.e., the equivalent of a point source, and a baseline level.

Although the assumptions that the bright spots are in fact unresolved
and that they can be separated in this way are not unquestionable, we
feel that this procedure gives a reasonable approximation of the
relative flux density of the two spots at each epoch, and is more
accurate than, for example, relying only the spot's peak brightness.

In Figure~\ref{ffspot} we show the 5-GHz evolution of the flux density
of each bright spot relative to the total flux density in the image as
a function of time.  The northeast bright spot's relative brightness
increases, reaching a maximum of $\gtrsim$20\% of the total flux
density at an age of $\sim$16~yr, and fades thereafter.  Recall,
however, that the total flux density is decreasing relatively rapidly
as a function of time (\S~\ref{svla}, Figure~\ref{flightcurve}), so in
absolute terms, the brightness of the northeast bright spot decreases
monotonically. At its relative peak, its flux density was $\sim
1.3$~mJy, and its radio luminosity, assuming an $\alpha = -0.61$
spectrum up to 100~GHz, was $\sim 6\times 10^{36}$~erg~s$^{-1}$.

The central bright spot was not clearly visible (at 5~GHz) till age
22.6~yr, but by age 25.6~yr, it had increased to $\sim$15\% of the
total flux density.  The central bright spot is still increasing in
brightness in absolute terms, but the absolute increase is not as
rapid as the relative one plotted in Figure~\ref{ffspot}.  In 2008,
the central bright spot had a flux density of $\sim 390 \;\mu$Jy, and
a radio luminosity of $\sim 2\times 10^{36}$~erg~s$^{-1}$ (again
assuming a spectrum with $\alpha = -0.61$ up to 100~GHz).

\section{DISCUSSION}
\subsection{Overall Evolution: Expansion Curve, Deceleration, and Spectral Index}
\label{sexpdecel}

The expansion curve in Figure~\ref{fexp} is compatible with a power-law
expansion, and earlier we found the best fit to correspond to $r \propto
t^{0.69}$.  The flux density decay, however, shows a distinct change
in the power-law slope at some time between 1989 and 1995, i.e.,
between ages of 6 to 12~yr \citep{SN86J-1}.  Unfortunately no flux
density measurements are available in this period to determine the
time of change more accurately.  Similar behavior was found for
SN~1993J with a steepening of the flux density decay at $t \sim$7~yr
\citep{SN93J-2, Weiler+2007}. 
In the case of SN~1993J, this steepening was found to correspond to a
decrease in deceleration \citep{SN93J-2}. This is the general
behavior expected when there is a relative decrease in the CSM 
density.  Is a similar behavior seen in SN~1986J?  Examining the
expansion curve in Figure~\ref{fexp} \citep[see also expansion curves
including earlier points in][]{SN86J-1, SN86J-COSPAR} suggests that a
significant {\em decrease}\/ in deceleration in the age range of 6 to
12~yr is not compatible with the data, although a small {\em increase}\/
in deceleration might be.  This would suggest that the increase in the
rate of flux density decay is due not to a relative decrease in the
CSM density but rather to one in the ejecta density.  This suggests a
steep density profile in the ejecta in the exterior regions, with a
somewhat flatter profile in the interior.  Such a flattening is indeed
expected from models of supergiant stars \citep[see, e.g., discussion
in][]{Chevalier2005}.  Although suggestive, we do not regard this
result as definitive because the measurements do not demand any break
in the power-law expansion curve.

The evolution of the integrated spectrum is complex, but at low
frequencies the spectrum seems to have a power-law shape. One can with
reasonable confidence associate the part of the spectrum at the lowest
frequencies with the shell, with the complex evolution due to the
emergence of the central bright spot being restricted to higher
frequencies ($\gtrsim 5$~GHz).  As we showed in \S~\ref{sspix} above,
the lowest part of the spectrum seems be initially rather steep, with
$\alpha = -0.67$, then by 1995 it flattens somewhat to $\alpha =
-0.44$, and subsequently it steepens again.

What could cause such a variation of the radio emission spectrum with
time?  Since the emission spectrum of synchrotron radiation reflects
the energy spectrum of the relativistic particles, one obvious
interpretation is that the latter energy spectrum is also changing
with time, perhaps in response to varying conditions at the shock
front where the particle acceleration occurs.  However, optical depth
effects also offer a possible explanation.  Although most of the
radio-emitting shell is optically thin at cm wavelengths, the interior
of the shell is not expected to be so because of the high density of
thermal material.  If mixing is occurring, a substantial transition
zone between the optically thin shocked material and the optically
thick ejecta may exist, which could partly absorb the radio emission
from a substantial fraction of the shell, and thus modify the
integrated spectrum.  In fact, the slow turn-on of the lightcurve
suggests a significant amount of absorption mixed in with the radio
emission already at early times \citep{WeilerPS1990}.

We note that changes with time in the radio spectral index are seen in
other supernovae.  In particular, a flattening with time of the radio
spectrum as seen for SN~1993J
\citep{SN93J-2}\footnote{\citet{ChandraRB2004b} also find a flattening
with time of the optically-thin spectrum of SN~1993J\@. A slight but
systematic flattening is also visible in the plotted data after $t =
500$~d of \citet{Weiler+2007}, although they fitted a model with a
constant spectral index for the optically thin part of the spectrum.}
and at least at the latest times also for SN~1979C
\citep{SN79C-shell}, may be characteristic of the evolution of radio
supernovae, and could be coupled with a steepening of the lightcurve
decay.  

\subsection{Comparison of Expansion Curve to Optical Radial Velocities}
\label{soptvel}

From our VLBI measurements, we determined the expansion velocity of
the edge of the radio emission region in the plane of the sky.  Since
the forward shock forms the outer boundary of the region where radio
emission is expected to occur, our VLBI expansion velocity can be
reasonably associated with the velocity of the forward shock velocity
\citep[see discussion on this subject for SN~1993J in][]{SN93J-4}.
Note that since the geometry of SN~1986J is somewhat irregular, the
assumption of sphericity which allows a comparison of the velocities
in the sky-plane measured with VLBI with those along the line-of-sight
measured from optical spectra is likely true only to first order.

In the optical, there are a number of spectral lines present which
allow radial velocity measurements.  Recently,
\citet{Milisavljevic+2008}, obtained new optical spectra at epoch
2007.7, and compared them to earlier ones from 1989.7 and 1991.8.
The optical emission lines are thought to mainly come from two
distinct locations.  Narrow optical lines, chiefly H$\alpha$, He~I and
[N~II] $\lambda 5755$, are thought to originate in dense, shock-heated
circumstellar medium, and therefore have relatively narrow
line-widths, typically $<2000$~\kms.  These lines were more prominent
early on and have faded since 1989.

Currently the most prominent emission lines are the [O I] $\lambda
\lambda 6300, 6364$ and [O II] $\lambda \lambda$ 7319, 7330 forbidden
lines of oxygen, with the [O III] $\lambda \lambda 4959, 5007$ line
also being discernible.  In 1989.7, these lines showed a double-peaked
structure, with one peak at a velocity of $-1000$~\kms\ (relative to
the systemic one), and the other near $-3500$~\kms, and the maximum
velocity, on the blue sides of these lines, was $\sim -6000$~\kms.
Similar velocities were seen in 1991.8\@.  By 2007.7, the double
structure was less clearly discernible, the maximum blue velocity had
decreased and the [O III] line had almost vanished.
\citet{Milisavljevic+2008} attribute these forbidden lines to ejecta
which has been heated (but not yet shocked) by the reverse shock.

This interpretation of the optical lines is consistent with the
expansion velocities measured in the radio.  In 1989.7, the forward
shock velocity was $\sim$8300~\kms, and the reverse shock velocity
would be expected to be about 80\% of this, or $\sim$6600~\kms.
The velocities implied by both the line-center and the total widths
seen in the forbidden oxygen lines are consistent with being inside
the reverse shock in 1989.7.  By 2007.7, the forward shock velocity is
$\sim$5500~\kms, and the expected reverse shock velocity
$\sim$4400~\kms.  The evolution of the optical line profiles is
consistent with this reduction in the velocity of the reverse shock in
that, particularly for the [O II] line, with the highest (blue)
velocities now being $\sim -4300$~\kms.  This suggests that, of the
ejecta responsible for the oxygen forbidden line emission, the highest
velocity fraction passed through the reverse shock between 1989.7 and
2007.7, and the remainder is now situated relatively close to the
reverse shock.

\subsection{Magnetic Field}
\label{sbfield}

The magnetic field can be estimated from the radio brightness and the
source size using minimum energy arguments
\citep{Pacholczyk1970}\footnote{We note that \citet{BeckK2005} point
out a number of shortcomings in the standard formulae of Pacholczyk
and give revised formulae.  However, for relatively flat spectra, the
differences between the standard and revised formulae are not large
and only weakly dependent on the field strength, so a field calculated
using the revised formula would have almost the same dependence on
time and radius.  In the interests of easier comparison with earlier
results we estimate our magnetic field using the standard formula from
\citet{Pacholczyk1970}.}.
Given that the spectrum shows deviations from a power-law, we will
perform the calculation only for the shell component, which dominates
below $\nu = 5$~GHz and is observed to have a power-law spectrum with
an average $\alpha = -0.5$ (\S~\ref{sspix}).  The derived field
strength depends on the ratio, $k$, of energies of relativistic
protons and electrons.  We use a value of $k = 100$, as was found to
be reasonable for strong shocks in supernova remnants by
\citet{BeckK2005}.  We estimate first the field strength in the shell,
for which we use the angular radii derived in \S~\ref{sexp} above, and
we take the flux density at 1.4~GHz to be representative of the shell
(as opposed to that of the central bright spot).

Using the values from 1988 to 2008, we find that $B$ has an
approximate power-law dependence on time and radius, as might be
expected from the approximately power-law evolution of both the flux
density and radius.  In particular, we find that for minimum energy,
$B \simeq 60 \, (t/{\rm 10 \, yr})^{-1.3}$~mG,
and that $B \propto r^{-1.8}$ for $t > 1500$~d. Although such
estimates of the field are uncertain by factors of several, the
estimates of its dependence on $t$ and $r$ are likely to be somewhat
better constrained with the exponents being uncertain by perhaps
$\sim$70\%.

\subsection{The Mass Loss Rate of the Progenitor}
\label{smassloss}

Can we estimate the mass-loss rate of the progenitor from our
observations?  This mass-loss rate is often estimated from studies of
the radio lightcurves.  It is determined mostly by the rising part of
the lightcurve, while the emission is optically thick, and is usually
expressed as $\dot M$/\vw, where \vw\ is the wind velocity.  However,
such calculations can significantly overestimate the value of $\dot
M$/\vw\ \citep[see, e.g.,][]{SN79C}, and in general have been shown to
be somewhat problematic if not based on complete physical models
\citep[e.g.,][]{FranssonB2005}.  In particular, in the case of
SN~1986J, \citet{WeilerPS1990} cited a value of $\dot M$/\vw\ =
2.4\ex{-4} \Msolxyr\ (for \vw\ = 10~\kms), based not on the rising
part of the lightcurve, but rather on a short term variation in
the optical depth seen in late 1988\@.  Later, \citet{Weiler+2002}
reported a notably lower value of 4.3\ex{-5} \Msolxyr, which is based
on a model involving clumpy absorption in the CSM\@.
\citet{Houck+1998} suggest a value of a few times $10^{-4}$~\Msolxyr\
based on X-ray observations.  In any case, given the departures from
self-similarity and the considerable evidence of clumpiness in the CSM,
which will have a strong effect on both radio brightness and
absorption, any estimated mass-loss rate is likely to be quite
uncertain.

A mass-loss rate between 4\ex{-5} and, say, 3\ex{-4} \Msolxyr\ with \vw
= 10~\kms\ would imply that by age 25.6~yr (2008.8), when the radius
of the shock-front is 6.3\ex{17}~cm, the swept-up mass would be $0.8
\sim 6$~\Msol.
For this range in mass-loss rate, the kinetic energy of the swept-up
mass would be $(0.2 \sim 1.7) \times 10^{51}$~erg ($v \simeq
5400$~\kms).  Since the kinetic energy of the swept-up mass can be at
most a fraction of the original kinetic energy of the ejecta, which
was likely not much larger than $10^{51}$~erg
\citep[e.g.,][]{Jones+1998}, we regard the higher end of this
mass-loss range as unlikely on energetic grounds.  We therefore
estimate a present swept-up mass of $0.8 \sim 2$~\Msol, and an average
mass-loss rate ($\dot M / v_{\rm w}$) in the range of $(4 \sim 10)
\times 10^{-5}$~\Msolxyr.
Such a range of mass-loss rate implies a CSM density at the 2008.8
outer shock radius of 6.3\ex{17}~cm               
of $(5 \sim 13)\ex{-22}$~g~cm$^{-3}$.  If we assume that the fraction
of hydrogen atoms is 75\% of the total, then the hydrogen number
density is between $230 \sim 570$~cm$^{-3}$.

\subsection{The Northeast Bright Spot}
\label{sshellspot}

The images are characterized by two bright spots, whose flux densities
vary.  The first one, in the northeast, appears to be a component of
the shell.  It was first clearly discernible at age $\sim$16~yr,
although it may have been present already earlier.  Its proper motion
is consistent with homologous expansion with the bulk of the shell
(Fig.~\ref{fvspot}).  As can be seen in Figure~\ref{ffspot}, at its
peak relative brightness (age $\sim$16~yr) approximately $\sim$20\% of
the total flux density originates from this spot.  Unfortunately, our
observations do not constrain the relative lightcurve around the time
of the peak precisely, the peak relative brightness may have been as
high as $\sim$30\% of the total, although it must have occurred
between our 1990 and 1999 epochs.  The northeast bright spot appears
marginally resolved in our images.  On 2005 Oct.\ 24 our FWHM resolution
was $1.65 \times 0.83$~mas, so the spot's angular diameter must have been 
$\sim$1~mas,
corresponding to 14\% of the shell's outer diameter, or 1.5\ex{17}~cm.

Since the northeast bright spot appears to be part of the projected
shell, the most plausible interpretation is that it is due to the
ejecta interacting with a dense clump in the CSM\@.  This would
account for the relative brightness of the spot first increasing and
then falling off.  In fact, the relative lightcurve of the northeast
bright spot in Figure~\ref{ffspot} suggests that the shock has
traversed the bulk of the clump between ages 10 to 20~yr.  Our
expansion curve suggests that, on average around the circumference,
the shock travelled a distance of 2\ex{17}~cm during this time, in
reasonable agreement with the clump diameter estimated above. 

If we assume both the dense clump and the outer shock to be spherical,
with the clump having the dimensions estimated above, then, when the
shock is midway through the clump, $\sim$1\% of the outer shock front
area is interacting with the dense clump.  Since the radio emission is
driven by the shock, we will consider the radio luminosity per unit
area of the shock, and we will further assume that the radio
luminosity scales as the 5-GHz flux density.
The northeast bright spot at its peak emitted about one quarter of the
5-GHz flux density of the remainder of the shell. we can therefore
calculate that the region in the dense clump must be $\sim$25 times
brighter than the remainder.  Our proper motion measurements show that
the velocity of the shock in the clump is approximately the same as it
is for the remainder of the shell, so the radio luminosity per unit
area of the shock should scale with the CSM density\footnote{The radio
luminosity is thought to scale with the post-shock energy density
\citep[e.g.,][]{Chevalier1998, Chevalier1982b}.  In the absence of
absorption, the radio luminosity will depend on the fractions of the
post-shock energy which go into the magnetic field and into
accelerated particles.  There is no solid ground for thinking that
these fractions are constant, indeed they are likely to vary somewhat
both as a function of shock velocity and the density and chemical
composition of the CSM\@.  It is likely, however, that these
variations are relatively small, whereas the variation in the CSM
density can be several orders of magnitude or more.}.
We found above (\S~\ref{smassloss}) that the present average CSM
density was $(5 \sim 13) \ex{-22}$~g~cm$^{-3}$, suggesting that the
densities in the clump are $(1.3 \sim 3.2) \ex{-20}$~g~cm$^{-3}$,
corresponding to hydrogen number densities of $5800 \sim
14,000$~cm$^{-3}$, giving the clump a total mass of $0.01 \sim
0.03$~\Msol.  We note, however, that these densities represent an
average over the clump, and much higher densities in parts of the
clump are not excluded by the radio data.

There are a number of other indicators pointing to the existence of
such dense clumps in the CSM of SN~1986J\@.  First, the very high
Balmer decrement \citep{Rupen+1987}, and other features in the
late-time spectra \citep{Milisavljevic+2008}, suggest the presence of
very high number densities ($> 10^6$~cm$^{-3}$).  Second, the slow
rise of the radio lightcurve is interpreted as also implying a very
clumpy CSM \citep{WeilerPS1990}. Third, because the H$\alpha$ line,
which has a width of only $\sim$1000~\kms\
\citep[e.g.,][]{Milisavljevic+2008, Leibundgut+1991}, is far too
narrow to arise from the ejecta near the forward shock ($v \sim
5700$~\kms; \S~\ref{sexp}), it has been suggested that most of the
H$\alpha$ emission is coming not from the ejecta, but rather from
shocked, dense clouds in the CSM resulting from a very clumpy wind
from the progenitor star \citep{ChugaiD1994, Chugai1993}.

Red supergiant stars of mass 20-40~\Msol\ are known to have episodes of
strong mass loss, resulting in large density contrasts in the CSM\@.
For example, \citet{Smith+2001, SmithHR2009} show that the red
supergiant VY Canis Majoris (mass 20-40~\Msol), which is likely
similar to the progenitor of SN~1986J, is surrounded by a nebula
produced by rapid mass loss ($2\sim4 \times 10^{-4}$~\Msolxyr).  This
nebula, which extends out to a radius of $\gtrsim 10^{17}$~cm, similar
in scale to the CSM required for SN~1986J, is highly structured with
knots and filamentary arcs, which are distributed in an asymmetric
way. \citet{Smith+2001} argue that the structures in the wind were
produced by ejection events which were localized on the stellar
surface.  For VY Canis Majoris then, and by implication also for
other red supergiants, the evidence seems to suggest that the winds
can be highly structured, with large density contrasts on small spatial
scales.  Exactly such structure seems to be required to reproduce the
radio emission from SN~1986J.

\subsection{The Central Bright Spot: Is It Emission from the Compact Remnant of
the Supernova Explosion?}
\label{scompact}

What is the nature of the central bright spot --- is it the emission
associated with a neutron-star or black-hole compact remnant of the
supernova explosion?  The first question would be whether the central
bright spot is positionally coincident with the explosion center.  As
we showed in \S~\ref{svspot} above, the position of the central bright
spot is in fact compatible with being in the center of the shell, and
its proper motion is consistent with being stationary with respect to
the shell.  Note that the expansion speed of the shell is much larger
than the peculiar velocities of stars and even of pulsars, which
latter have a velocity dispersion of $200 \sim 300$~\kms\ in the
Galaxy \citep[e.g.,][]{Hobbs+2005,LyneG1990},
so the effect of any peculiar motion of the progenitor or any ``kick''
received in the explosion would be small and the compact remnant would
be expected to remain essentially in the center of the shell over the
$\sim$25~yrs since the supernova explosion.  In this respect, our
position and proper motion measurements for the central hot spot are
consistent with the expectations for a black-hole or neutron star
compact remnant.

We first noted the inversion in the integrated radio spectrum in
\citet{SN86J-1}, and we showed in \citet{SN86J-Sci} that the inversion
in the integrated spectrum was in fact associated with the central
bright spot, with the shell and the northeast bright spot discussed
above having a simple power-law spectrum as is usually seen in
supernovae.  In \S~\ref{svla} and Figure~\ref{fspectra} above we show
that both the inflection and the high-frequency turnover points in the
radio spectrum are progressing downwards in frequency as SN~1986J
ages.  The inverted spectrum of the central bright spot suggests
partly absorbed radio emission. An opacity decreasing due to the
expansion would cause the observed progression to lower frequencies of
both the inversion and high-frequency turnover points.

We argued in \citet{SN86J-COSPAR} that the absorption is probably due
to free-free absorption rather than to synchrotron self-absorption
(SSA)\@.  In fact, our latest 22~GHz images, where the central bright
spot is somewhat resolved, largely rules out SSA on the following
grounds.  Examination of the image (Fig.~\ref{fimage22}b) suggests a
size of $\gtrsim 1 \times 10^{17}$~cm for the central bright spot.
The radius below which SSA would be important, $R_p$, can be
calculated following \citet{Chevalier1998}:

\[
R_p = 8.8\times10^{15}\kappa^{-1/10} 
 \left( \frac{f}{0.5}\right)^{-1/19} 
 \left( \frac{F_p}{\rm Jy}\right)^{9/19} \]
\[
 \times
 \left( \frac{D}{\rm Mpc}\right)^{18/19} 
 \left( \frac{\nu}{\rm 5 GHz}\right)^{-1} {\rm cm}
\]
where $F_p$ is the flux density at the spectral peak, $\kappa$ is the
ratio of relativistic electron energy density to magnetic energy
density, $D$ is the distance, and $f$ is the filling factor.  Assuming
equipartition ($\kappa = 1$) and substituting our values of $F_p$ =
3.8~mJy at $\nu_p$ = 14~GHz (2007.6), we obtain $R_p \simeq 1 \times
10^{16}$~cm,
which is far lower than the size determined from the image of $\gtrsim
1 \times 10^{17}$~cm, thus effectively ruling out significant SSA at
our observing frequencies.

The appearance of the central bright spot at a time long after the
supernova's integrated radio spectrum was optically thin at cm
wavelengths suggested that any remaining absorption due to the
circumstellar material was minimal.  It was natural, therefore, to
assume that the absorption seen for the central bright spot was due to
un-shocked material interior to the shocked shell, rather than to the
CSM\@.  With this assumption, the central bright spot's likeliest
physical location was in the center of the shell, and its most likely
interpretation was in terms of radio emission associated with the
neutron star or black hole expected to have been left behind after the
explosion \citep{SN86J-Sci, SN86J-COSPAR, SN86J-COSPAR-2}.  

The turnover frequency is approximately equal to
the frequency at which the optical depth to free-free absorption,
$\tau$, is unity, is given by
$$\nu_{\tau = 1} = 0.3 \, (T_e^{-1.35} N_e^2 \, dl)^{1/2} \; {\rm
GHz,}$$ where $T_e$ is the electron temperature in K, $N_e$ is the
number density of electrons in cm$^{-3}$, assumed constant along
the path-length in pc, $dl$, with $N_e^2 \, dl$ being the emission
measure.
Our integrated spectra (Fig.~\ref{fspectra}) show that the spectral
peak due to the central bright spot occurs at $\sim$20~GHz in 2002.4.
If we assume\footnote{Although $T_e$ is not well
known, and can vary strongly with radius, this is probably a
reasonable value for radii $>1 \times 10^{17}$~cm \citep[see,
e.g.,][]{LundqvistF1988}.},  a $T_e$ of $10^4$~K then $N_e^2 \, dl$ is
$1.1\ex{9}$~cm$^{-6}$~pc.  By 2007.6, the turnover frequency is
$\sim$14~GHz, so $N_e^2 \, dl = 0.6\ex{9}$~cm$^{-6}$~pc.

The temporal dependence of the absorption is consistent with that
expected from the gas within the expanding shell.  We found that the
emission measure decreased by a factor of $\sim$2 between 2002.4 and
2007.6.  If we assume the system is expanding homologously, then the
density within the shell is proportional to $r^{-3}$ and $N_e^2 \, dl
\propto r^{-5}$.  If we further take $r \propto t^{0.69}$ as we found
for the shell, we would expect $N_e^2 \, dl$ to be $\propto
t^{-3.45}$, which would lead to a decrease by close to the observed
factor of 2 between 2002.4 and 2007.6.

We note, however, that the inverted part of the spectrum is much
flatter than is expected from free-free absorption, which produces
spectra with an exponential cutoff at low frequencies.  Such flatter
spectra below the turnover are produced when a range of different
optical depths is present.  This suggests that the absorbing material
is somewhat fragmented, with some lines of sight having much lower
optical depths than others.  Indeed, fragmentation of the ejecta is
not surprising, since there are a number of instabilities operating in
the expanding shell of ejecta
\citep[e.g.,][]{Gull1973,JunN1996a,ChevalierB1995}, which are expected
to lead to fragmentation.  If the central bright spot is due to a
pulsar, then further instabilities occur when the young pulsar's wind
nebula expands into the supernova ejecta
\citep[e.g.,][]{BandieraPS1983,ChevalierF1992}.

As we have shown, the absorption of the radio emission from the
central bright spot is consistent with what might be expected from the
intervening material in the shell and its observed expansion rate.
This suggests that the integrated spectrum above the turnover point
shows the intrinsic, un-absorbed spectrum of the central bright spot.
In fact, the integrated spectra in Figure~\ref{fspectra} show that the
spectral index above the turnover point is very similar to that below
the inversion point.  So, the intrinsic spectral index of the central
bright spot is also similar to that of the shell.

If the central bright spot were in fact in the physical center of the
shell and represented radio emission associated with the compact
remnant of the supernova --- either a black hole or a neutron star ---
then it would be somewhat of a coincidence for its radio emission to
have a similar spectral index as that of the shell emission.  In
particular, if the central bright spot were the wind nebula (PWN)
around a very young pulsar, then one might expect it to have a
somewhat flatter spectral index, as most PWNe have spectral indices in
the range $-0.3$ to 0.0 \citep{GaenslerS2006} and as there are no
filled-center remnants in the catalog of \citet{Green2004} which have
broadband radio spectra steeper\footnote{see {\tt
http://www.mrao.cam.ac.uk/surveys/snrs} for an updated version of the
catalog.  The possible exception is Vela~X, the central source in the
Vela supernova remnant, which is often associated with the PWN\@.
\citet{Alvarez+2001} found $\alpha^{\rm 8.4 GHz}_{\rm 0.09 GHz} =
-0.39 \pm 0.03$ for Vela~X, indicating a relatively steep spectrum for
a PWN\@. More recently, however, \citet{Hales+2004} reported on 31~GHz
observations of a strong source near the Vela pulsar that they
identify with the PWN, and for which they find a slightly inverted
spectrum with $\alpha^{\rm 31 GHz}_{\rm 8.4 GHz} = +0.10 \pm 0.06$}
than $\alpha = -0.3$.  In particular, the youngest known PWNe, the
Crab Nebula and G21.5$-$0.5, have rather flat radio spectra with
spectral indices of $-0.27$ \citep{Crab-1997} and +0.08
\citep{G21.5expand}, respectively.  The observed steep spectrum of the
central bright spot, therefore, would not seem to favor the hypothesis
of a young PWN\@.  However, an only 27-yr old PWN may not be comparable
to the much older known PWNe, so the observed steep spectrum does
not rule out a young PWN.

Alternatively, if the central bright spot represents emission from
jets emanating from a black hole environment, the radio spectrum is
consistent with the expectations but the radio luminosity is not.  The
possible presence of both self-absorbed and optically thin components
in the jet can give rise to both flat and steep spectra, so the
observed spectrum can easily be accommodated.  However, SN~1986J's
central bright spot is far more radio-luminous than any known
stellar-mass black hole system.  In particular, a ``fundamental
plane'' relationship between the radio and X-ray luminosities and
black-hole mass has been observed for black-hole systems of a wide
range of masses \citep[see, e.g.,][]{FalckeKM2004, Ho2008}.  The
unabsorbed X-ray flux from SN~1986J was measured in late 2003 by
\citet{Houck2005a} to be $\sim 1.6\times 10^{-13}$~erg~s$^{-1}$.  Even
with the assumption that all this X-ray flux is due to the central
bright spot, the ``fundamental plane'' relation would suggest radio
luminosities for a black-hole system of $\sim 10^{-4}$ of that
observed for the central bright spot of $\sim 2 \times
10^{36}$~erg~s$^{-1}$ (\S~\ref{sfspot}).  However, very little is
known about such young black-hole systems, so it is not impossible
that they would be far more luminous than expected from the
fundamental plane relationship.  So again, we must conclude that the
radio properties of the central bright spot do not seem to argue
strongly for an accreting black-hole system, but neither can they
exclude it.

\subsection{The Central Bright Spot as a Shell Component?}
\label{scenterclump}

As we showed in \S~\ref{sshellspot} above, the northeast bright spot
can naturally be explained as the impact of the shock-front on a dense
condensation in the CSM\@.  Could the central bright spot be a similar
phenomenon, with the dense condensation being coincidentally located
near the projected center of the shell?  In this case, an approximate
equality of the intrinsic spectral indices of the central bright spot
with that of the remainder of the supernova is expected, since usually
not much variation of the spectral index around the shell is observed.

As we showed, there is considerable evidence that the CSM of red
supergiants and of SN~1986J in particular are quite clumpy, and the
northeast bright spot suggests that there is at least one dense
condensation in the CSM of SN~1986J\@. There is no particular reason
to only expect a single dense clump in the CSM, so the existence of
more than one dense clump seems not unlikely.

The expanding shell impacting on a dense CSM clump would cause a local
brightening of the radio emission.  Although the bulk of the CSM is
optically thin at cm wavelengths, clumps with densities as high as
$N_e > 10^6$~cm$^{-3}$, and sizes small compared to the shell
diameter, would be expected to have large optical depths for cm-wave
radio emission.  For example, a clump with a diameter of 5\ex{16}~cm,
corresponding to an angle of $\sim$0.3~mas on our images, and a
density of 1\ex{6}~cm$^{-3}$ would have
a free-free optical depth at 8.4~GHz of $\sim$30.

The shock hitting a dense clump will produce localized, bright radio
emission, as is seen in the northeast bright spot.  However, if the
dense CSM clump lies on the near side of the expanding shell,
fortuitously near the center in projection, then the optically thick
clump itself will in fact block most of this bright radio emission.
With time, the clump will become optically thin as it either fragments
or as the shock eats through it.  This scenario would produce a
sequence much like what is seen for the central bright spot: a delayed
turn on, dependent on the distance of the clump from the explosion
center and its density, with an inverted spectrum expected from
free-free absorption, followed eventually by an optically thin decay.
At 5~GHz, the brightness of the central bright spot is still
increasing, (Fig.~\ref{ffspot}) so we are still in the rising part of
its lightcurve.

If we interpret the central bright spot as emission from the shock
interacting with such a CSM clump, what can be deduced about the
physical conditions in this clump?  In our latest image at 22~GHz, we
can estimate an angular diameter of the central bright spot of
$\sim$0.7~mas (FWHM), suggesting a linear diameter of $\sim 1 \times
10^{17}$~cm or 0.03~pc.  The turnover frequency at this
epoch is $\sim$12~GHz.  If we take the turnover frequency to be that
at which the free-free optical depth is unity, the density of the
absorbing clump can be estimated from the electron number density
which in turn can be calculated from the emission measure using the
equation given in \S~\ref{scompact} above. If we again assume $T_e =
10^4$~K 
and a constant $N_e$, and further assume that the
clump's line-of-sight depth is equal to the central bright spot's
diameter in the plane of the sky, we can calculate that $N_e \sim
2.5\ex{5}$~cm$^{-3}$.  Again, the presence of
significantly higher densities cannot be excluded, but the presence of
substantial material at the calculated density seems required.  The
total mass of such an absorbing clump would be $\sim$0.1~\Msol\
(assuming full ionization). 
The swept-up mass in 2008.8 is $0.8 \sim 2.0$~\Msol\ (out to a radius of
$6.3\times10^{17}$~cm; see \S~\ref{smassloss}), so the mass of the
clump represents 13\% to 5\% of the total swept-up mass. 

By comparing the part of the spectrum above 20~GHz with that at low
frequencies (see Fig.~\ref{fspectra}), we can estimate that the
unabsorbed radio luminosity of the central bright spot is about twice
that
of the remainder of the supernova. Note that the unabsorbed luminosity
of the central bright spot is currently larger than that of the
northeast bright spot, since a substantial fraction of the central
bright spot's luminosity is still absorbed, with the turnover
frequency currently being $\sim$12~GHz.  If we again assume that radio
luminosity per unit shock area scales only with the CSM density, we
can estimate the density of the clump relative to that of the average
CSM density as we did for the northeast bright spot
(\S~\ref{sshellspot}).  Using the above clump diameter of 1\ex{17}~cm,
and a shell outer radius of $6\ex{17}$~cm (for 2006 Dec.)\ we find
that the spot covers
$\sim$0.2\% of the shell surface.  The clump must therefore be $\sim
1000\times$ denser than the average corresponding CSM\@.  As might be
expected, since the central bright spot has a much higher unabsorbed
luminosity than the northeast bright spot, the density required for
the relevant enhancement of the emissivity is also much higher.

We calculated a representative density of $N_e \sim
2.5\ex{5}$~cm$^{-3}$ for the central clump above, which suggests a
representative average density for the CSM at $r \simeq 6.2\ex{17}$~cm
of $\sim 250$~cm$^{-3}$, which 
in turn implies a mass-loss rate of 6\ex{-5} \Msolxyr\ (for \vw =
10~\kms) which is consistent with the values derived in
\S~\ref{smassloss} above.

As mentioned in \S~\ref{soptvel} above, the oxygen forbidden-line
emission seems to be dominated by two dense clumps, both on the near
side of the remnant.  If the central radio bright spot is due to a CSM
clump on the near side of the SN, then also in the radio, there would
be evidence for two dense clumps.  Could the clumps responsible for
radio bright spots be be the same as the two clumps seen in the
optical emission lines?  The association seems unclear, since the
radio emission, as we have argued, would be due to dense clumps in the
CSM, exterior to the forward shock, while the forbidden-line emission
is plausibly attributed ejecta heated by the reverse shock, that is
interior to the reverse shock.

While it is conceivable that dense CSM clumps cause distortions first
in the forward shock but subsequently also in the reverse shock, which
latter could in turn give rise to the bright features seen in the
oxygen forbidden lines, any association between the radio bright spots
and the oxygen-line features must remain speculative.


All said, we think now that the spectral evolution seen in the new
observations make an interpretation of the central bright spot in
terms of a second shell component and an interpretation in terms of
emission from a PWN or a black-hole environment equally plausible.

\section{SUMMARY AND CONCLUSIONS}

\begin{trivlist}

\item{1.} We have obtained new multi-frequency VLA flux density
measurements and VLBI images of SN~1986J, showing the evolution of this
supernova in the radio.

\item{2.} The evolution of the integrated radio spectrum is complex.
The spectrum at the lowest frequencies has a spectral index in the
range of $-0.7 \sim -0.5$.  The spectrum above the high-frequency
turnover, presumably the intrinsic spectrum of the central bright spot,
is equally steep within the uncertainties.

\item{3.} The shell continues to expand.  The average expansion speed
of the shell between 1999 and 2009 was $5700 \pm 1000$~\kms.  This
speed is compatible with continued power-law expansion, with the
radius increasing $\propto t^{0.69\pm0.03}$.  The increase in the rate
of flux density decay at $t \sim 7$~yr is likely due to a flattening
in the profile of the ejecta profile rather than a steepening in the
one of the CSM, as no increase in deceleration is observed.

\item{4.} The equipartition magnetic field decreases as the supernova
expands, with an approximate value of $B \simeq 60 \; (t/{\rm
10 \; yr})^{-1.3}$~mG\@.  The dependence on the outer shock front radius is
$B \propto r^{-1.8}$.

\item{5.} Various estimates of the mass-loss converge to values in the
range of $(4 \sim 10) \times 10^{-5}$ \Msolxyr\ (for \vw = 10~\kms).

\item{6.} The VLBI images show two bright spots in addition to the
shell-like structure.  The first is in the northeast of the shell,
while the second is near the projected center of the shell.  The flux
densities of the spots relative to that of the SN as a whole vary with
time, with that of the northeast bright spot increasing till
$\sim$1999, and decreasing since, while that of the central bright
spot continues to increase.

\item{7.} The northeast bright spot is likely due to a dense clump in
the circumstellar material (CSM).  The bright spot's proper motion is
consistent with homologous power-law expansion.  Number densities of
$N_e \sim 10^4$~cm$^{-3}$ or higher are suggested for the clump.

\item{8.} The central bright spot has a partly absorbed radio
spectrum, with an intrinsic radio spectrum that is similar to that of
the shell.  The amount of absorption is decreasing with time.

\item{9.} The central bright spot's original interpretation as
originating in the physical center of the shell and being emission due
to the neutron star or black hole remnant of the supernova is
supported by its central position, and its stationarity to within
$1\sigma$.  The new observations, however, suggest an equally
plausible alternative explanation of the central bright spot being
radio emission due to the shell impacting upon a second dense CSM
clump, fortuitously located on the near side of the shell close to the
projected center.  The amount of absorption suggests number densities
of $N_e \gtrsim 2.5$\ex{5}~cm$^{-3}, \sim 10^3$~higher than
the average density of the CSM.

\item{10.} From the VLBI images from 1987 to 2008, we produced a movie
showing the supernova's evolution.

\end{trivlist}

\acknowledgements 

The European VLBI Network is a joint facility of European and Chinese
radio astronomy institutes funded by their national research
councils. We have made use of NASA's Astrophysics Data System Abstract
Service. We thank the anonymous referee for his suggestions.

\bibliographystyle{apj}
\bibliography{mybib1}

\begin{thebibliography}{62}
\expandafter\ifx\csname natexlab\endcsname\relax\def\natexlab#1{#1}\fi

\bibitem[{{Aaronson} {et~al.}(1982){Aaronson}, {Huchra}, {Mould}, {Tully},
  {Fisher}, {van Woerden}, {Goss}, {Chamaraux}, {Mebold}, {Siegman},
  {Berriman}, \& {Persson}}]{Aaronson+1982}
{Aaronson}, M., {et~al.} 1982, \apjs, 50, 241

\bibitem[{{Alvarez} {et~al.}(2001){Alvarez}, {Aparici}, {May}, \&
  {Reich}}]{Alvarez+2001}
{Alvarez}, H., {Aparici}, J., {May}, J., \& {Reich}, P. 2001, \aap, 372, 636

\bibitem[{{Baars} {et~al.}(1977){Baars}, {Genzel}, {Pauliny-Toth}, \&
  {Witzel}}]{Baars+1977}
{Baars}, J.~W.~M., {Genzel}, R., {Pauliny-Toth}, I.~I.~K., \& {Witzel}, A.
  1977, \aap, 61, 99

\bibitem[{{Ball} \& {Kirk}(1995)}]{BallK1995}
{Ball}, L., \& {Kirk}, J.~G. 1995, \aap, 303, L57

\bibitem[{{Bandiera} {et~al.}(1983){Bandiera}, {Pacini}, \&
  {Salvati}}]{BandieraPS1983}
{Bandiera}, R., {Pacini}, F., \& {Salvati}, M. 1983, \aap, 126, 7

\bibitem[{{Bartel} \& {Bietenholz}(2003)}]{SN79C}
{Bartel}, N., \& {Bietenholz}, M.~F. 2003, \apj, 591, 301

\bibitem[{{Bartel} \& {Bietenholz}(2008)}]{SN79C-shell}
---. 2008, \apj, 682, 1065

\bibitem[{{Bartel} {et~al.}(2007){Bartel}, {Bietenholz}, {Rupen}, \&
  {Dwarkadas}}]{SN93J-4}
{Bartel}, N., {Bietenholz}, M.~F., {Rupen}, M.~P., \& {Dwarkadas}, V.~V. 2007,
  \apj, 668, 924

\bibitem[{{Bartel} {et~al.}(1991){Bartel}, {Rupen}, {Shapiro}, {Preston}, \&
  {Rius}}]{Bartel+1991}
{Bartel}, N., {Rupen}, M.~P., {Shapiro}, I.~I., {Preston}, R.~A., \& {Rius}, A.
  1991, \nat, 350, 212

\bibitem[{{Bartel} {et~al.}(1987){Bartel}, {Ratner}, {Rogers}, {Shapiro},
  {Bonometti}, {Cohen}, {Gorenstein}, {Marcaide}, \& {Preston}}]{Bartel+1987}
{Bartel}, N., {et~al.} 1987, \apj, 323, 505

\bibitem[{{Bartel} {et~al.}(2002){Bartel}, {Bietenholz}, {Rupen}, {Beasley},
  {Graham}, {Altunin}, {Venturi}, {Umana}, {Cannon}, \& {Conway}}]{SN93J-2}
---. 2002, \apj, 581, 404

\bibitem[{{Beck} \& {Krause}(2005)}]{BeckK2005}
{Beck}, R., \& {Krause}, M. 2005, Astronomische Nachrichten, 326, 414

\bibitem[{{Bietenholz} \& {Bartel}(2008{\natexlab{a}})}]{SN86J-COSPAR-2}
{Bietenholz}, M.~F., \& {Bartel}, N. 2008{\natexlab{a}}, Advances in Space
  Research, 41, 424

\bibitem[{{Bietenholz} \& {Bartel}(2008{\natexlab{b}})}]{G21.5expand}
---. 2008{\natexlab{b}}, \mnras, 386, 1411

\bibitem[{{Bietenholz} {et~al.}(2000){Bietenholz}, {Bartel}, \&
  {Rupen}}]{M81-2000}
{Bietenholz}, M.~F., {Bartel}, N., \& {Rupen}, M.~P. 2000, \apj, 532, 895

\bibitem[{{Bietenholz} {et~al.}(2001){Bietenholz}, {Bartel}, \&
  {Rupen}}]{SN93J-1}
---. 2001, \apj, 557, 770

\bibitem[{{Bietenholz} {et~al.}(2002){Bietenholz}, {Bartel}, \&
  {Rupen}}]{SN86J-1}
---. 2002, \apj, 581, 1132

\bibitem[{{Bietenholz} {et~al.}(2004{\natexlab{a}}){Bietenholz}, {Bartel}, \&
  {Rupen}}]{SN86J-Sci}
---. 2004{\natexlab{a}}, Science, 304, 1947

\bibitem[{{Bietenholz} {et~al.}(2004{\natexlab{b}}){Bietenholz}, {Bartel}, \&
  {Rupen}}]{M81-2004}
---. 2004{\natexlab{b}}, \apj, 615, 173

\bibitem[{{Bietenholz} {et~al.}(2005){Bietenholz}, {Bartel}, \&
  {Rupen}}]{SN86J-COSPAR}
---. 2005, Advances in Space Research, 35, 1052

\bibitem[{{Bietenholz} {et~al.}(1997){Bietenholz}, {Kassim}, {Frail}, {Perley},
  {Erickson}, \& {Hajian}}]{Crab-1997}
{Bietenholz}, M.~F., {Kassim}, N., {Frail}, D.~A., {Perley}, R.~A., {Erickson},
  W.~C., \& {Hajian}, A.~R. 1997, \apj, 490, 291

\bibitem[{{Chandra} {et~al.}(2004){Chandra}, {Ray}, \&
  {Bhatnagar}}]{ChandraRB2004b}
{Chandra}, P., {Ray}, A., \& {Bhatnagar}, S. 2004, \apj, 612, 974

\bibitem[{{Chevalier} \& {Blondin}(1995)}]{ChevalierB1995}
{Chevalier}, R., \& {Blondin}, J.~M. 1995, \apj, 444, 312

\bibitem[{{Chevalier}(1982)}]{Chevalier1982b}
{Chevalier}, R.~A. 1982, \apj, 259, 302

\bibitem[{{Chevalier}(1987)}]{Chevalier1987}
---. 1987, \nat, 329, 611

\bibitem[{{Chevalier}(1998)}]{Chevalier1998}
---. 1998, \apj, 499, 810

\bibitem[{{Chevalier}(2005)}]{Chevalier2005}
---. 2005, \apj, 619, 839

\bibitem[{{Chevalier} \& {Fransson}(1992)}]{ChevalierF1992}
{Chevalier}, R.~A., \& {Fransson}, C. 1992, \apj, 395, 540

\bibitem[{{Chugai}(1993)}]{Chugai1993}
{Chugai}, N.~N. 1993, \apjl, 414, L101

\bibitem[{{Chugai} \& {Danziger}(1994)}]{ChugaiD1994}
{Chugai}, N.~N., \& {Danziger}, I.~J. 1994, \mnras, 268, 173

\bibitem[{{Falcke} {et~al.}(2004){Falcke}, {K{\"o}rding}, \&
  {Markoff}}]{FalckeKM2004}
{Falcke}, H., {K{\"o}rding}, E., \& {Markoff}, S. 2004, \aap, 414, 895

\bibitem[{{Ferrarese} {et~al.}(2000){Ferrarese}, {Ford}, {Huchra}, {Kennicutt},
  {Mould}, {Sakai}, {Freedman}, {Stetson}, {Madore}, {Gibson}, {Graham},
  {Hughes}, {Illingworth}, {Kelson}, {Macri}, {Sebo}, \&
  {Silbermann}}]{Ferrarese+2000}
{Ferrarese}, L., {et~al.} 2000, \apjs, 128, 431

\bibitem[{{Fey} {et~al.}(2004){Fey}, {Ma}, {Arias}, {Charlot},
  {Feissel-Vernier}, {Gontier}, {Jacobs}, {Li}, \& {MacMillan}}]{Fey+2004}
{Fey}, A.~L., {et~al.} 2004, \aj, 127, 3587

\bibitem[{{Fransson} \& {Bj{\"o}rnsson}(2005)}]{FranssonB2005}
{Fransson}, C., \& {Bj{\"o}rnsson}, C.-I. 2005, in IAU Colloq. 192: Cosmic
  Explosions, On the 10th Anniversary of SN1993J, ed. J.-M. {Marcaide} \& K.~W.
  {Weiler}, 59

\bibitem[{{Gaensler} \& {Slane}(2006)}]{GaenslerS2006}
{Gaensler}, B.~M., \& {Slane}, P.~O. 2006, \araa, 44, 17

\bibitem[{{Green}(2004)}]{Green2004}
{Green}, D.~A. 2004, Bulletin of the Astronomical Society of India, 32, 335

\bibitem[{{Gull}(1973)}]{Gull1973}
{Gull}, S.~F. 1973, \mnras, 161, 47

\bibitem[{{Hales} {et~al.}(2004){Hales}, {Casassus}, {Alvarez}, {May},
  {Bronfman}, {Readhead}, {Pearson}, {Mason}, \& {Dodson}}]{Hales+2004}
{Hales}, A.~S., {et~al.} 2004, \apj, 613, 977

\bibitem[{{Ho}(2008)}]{Ho2008}
{Ho}, L.~C. 2008, \araa, 46, 475

\bibitem[{{Hobbs} {et~al.}(2005){Hobbs}, {Lorimer}, {Lyne}, \&
  {Kramer}}]{Hobbs+2005}
{Hobbs}, G., {Lorimer}, D.~R., {Lyne}, A.~G., \& {Kramer}, M. 2005, \mnras,
  360, 974

\bibitem[{{Houck}(2005)}]{Houck2005a}
{Houck}, J.~C. 2005, in X-Ray and Radio Connections (eds. L.O. Sjouwerman and
  K.K Dyer) Published electronically by NRAO,
  http://www.aoc.nrao.edu/events/xraydio Held 3-6 February 2004 in Santa Fe,
  New Mexico, USA, (E3.03) 6 pages

\bibitem[{{Houck} {et~al.}(1998){Houck}, {Bregman}, {Chevalier}, \&
  {Tomisaka}}]{Houck+1998}
{Houck}, J.~C., {Bregman}, J.~N., {Chevalier}, R.~A., \& {Tomisaka}, K. 1998,
  \apj, 493, 431

\bibitem[{{Jones} {et~al.}(1998){Jones}, {Rudnick}, {Jun}, {Borkowski},
  {Dubner}, {Frail}, {Kang}, {Kassim}, \& {McCray}}]{Jones+1998}
{Jones}, T.~W., {et~al.} 1998, \pasp, 110, 125

\bibitem[{{Jun} \& {Norman}(1996)}]{JunN1996a}
{Jun}, B., \& {Norman}, M.~L. 1996, \apj, 465, 800

\bibitem[{{Kovalev} {et~al.}(2008){Kovalev}, {Lobanov}, {Pushkarev}, \&
  {Zensus}}]{Kovalev+2008a}
{Kovalev}, Y.~Y., {Lobanov}, A.~P., {Pushkarev}, A.~B., \& {Zensus}, J.~A.
  2008, \aap, 483, 759

\bibitem[{{Kraan-Korteweg}(1986)}]{Kraan-Korteweg1986}
{Kraan-Korteweg}, R.~C. 1986, \aaps, 66, 255

\bibitem[{{Leibundgut} {et~al.}(1991){Leibundgut}, {Kirshner}, {Pinto},
  {Rupen}, {Smith}, {Gunn}, \& {Schneider}}]{Leibundgut+1991}
{Leibundgut}, B., {Kirshner}, R.~P., {Pinto}, P.~A., {Rupen}, M.~P., {Smith},
  R.~C., {Gunn}, J.~E., \& {Schneider}, D.~P. 1991, \apj, 372, 531

\bibitem[{{Lundqvist} \& {Fransson}(1988)}]{LundqvistF1988}
{Lundqvist}, P., \& {Fransson}, C. 1988, \aap, 192, 221

\bibitem[{{Lyne} \& {Graham-Smith}(1990)}]{LyneG1990}
{Lyne}, A.~G., \& {Graham-Smith}, F. 1990, {Pulsar astronomy}, ed. F.~Lyne, A.
  G. \& Graham-Smith

\bibitem[{{Milisavljevic} {et~al.}(2008){Milisavljevic}, {Fesen}, {Leibundgut},
  \& {Kirshner}}]{Milisavljevic+2008}
{Milisavljevic}, D., {Fesen}, R.~A., {Leibundgut}, B., \& {Kirshner}, R.~P.
  2008, \apj, 684, 1170

\bibitem[{{Pacholczyk}(1970)}]{Pacholczyk1970}
{Pacholczyk}, A.~G. 1970, {Radio astrophysics. Nonthermal processes in galactic
  and extragalactic sources} (San Francisco: Freeman)

\bibitem[{{P{\'e}rez-Torres} {et~al.}(2002){P{\'e}rez-Torres}, {Alberdi},
  {Marcaide}, {Guirado}, {Lara}, {Mantovani}, {Ros}, \&
  {Weiler}}]{Perez-Torres+2002}
{P{\'e}rez-Torres}, M.~A., {Alberdi}, A., {Marcaide}, J.~M., {Guirado}, J.~C.,
  {Lara}, L., {Mantovani}, F., {Ros}, E., \& {Weiler}, K.~W. 2002, \mnras, 335,
  L23

\bibitem[{{Pradel} {et~al.}(2006){Pradel}, {Charlot}, \&
  {Lestrade}}]{PradelCL2006}
{Pradel}, N., {Charlot}, P., \& {Lestrade}, J.-F. 2006, \aap, 452, 1099

\bibitem[{{Rupen} {et~al.}(1987){Rupen}, {van Gorkom}, {Knapp}, {Gunn}, \&
  {Schneider}}]{Rupen+1987}
{Rupen}, M.~P., {van Gorkom}, J.~H., {Knapp}, G.~R., {Gunn}, J.~E., \&
  {Schneider}, D.~P. 1987, \aj, 94, 61

\bibitem[{{Smith} {et~al.}(2009){Smith}, {Hinkle}, \& {Ryde}}]{SmithHR2009}
{Smith}, N., {Hinkle}, K.~H., \& {Ryde}, N. 2009, \aj, 137, 3558

\bibitem[{{Smith} {et~al.}(2001){Smith}, {Humphreys}, {Davidson}, {Gehrz},
  {Schuster}, \& {Krautter}}]{Smith+2001}
{Smith}, N., {Humphreys}, R.~M., {Davidson}, K., {Gehrz}, R.~D., {Schuster},
  M.~T., \& {Krautter}, J. 2001, \aj, 121, 1111

\bibitem[{{Tonry} {et~al.}(2001){Tonry}, {Dressler}, {Blakeslee}, {Ajhar},
  {Fletcher}, {Luppino}, {Metzger}, \& {Moore}}]{Tonry+2001}
{Tonry}, J.~L., {Dressler}, A., {Blakeslee}, J.~P., {Ajhar}, E.~A., {Fletcher},
  A.~B., {Luppino}, G.~A., {Metzger}, M.~R., \& {Moore}, C.~B. 2001, \apj, 546,
  681

\bibitem[{{Tully}(1988)}]{Tully1988}
{Tully}, R.~B. 1988, {Nearby galaxies catalog} (Cambridge and New York,
  Cambridge University Press, 1988, 221 p.)

\bibitem[{{van Gorkom} {et~al.}(1986){van Gorkom}, {Rupen}, {Knapp}, {Gunn},
  {Neugebauer}, \& {Matthews}}]{vGorkom+1986}
{van Gorkom}, J., {Rupen}, M., {Knapp}, G., {Gunn}, J., {Neugebauer}, G., \&
  {Matthews}, K. 1986, \iaucirc, 4248, 1

\bibitem[{{Weiler} {et~al.}(2002){Weiler}, {Panagia}, {Montes}, \&
  {Sramek}}]{Weiler+2002}
{Weiler}, K.~W., {Panagia}, N., {Montes}, M.~J., \& {Sramek}, R.~A. 2002,
  \araa, 40, 387

\bibitem[{{Weiler} {et~al.}(1990){Weiler}, {Panagia}, \&
  {Sramek}}]{WeilerPS1990}
{Weiler}, K.~W., {Panagia}, N., \& {Sramek}, R.~A. 1990, \apj, 364, 611

\bibitem[{{Weiler} {et~al.}(2007){Weiler}, {Williams}, {Panagia}, {Stockdale},
  {Kelley}, {Sramek}, {Van Dyk}, \& {Marcaide}}]{Weiler+2007}
{Weiler}, K.~W., {Williams}, C.~L., {Panagia}, N., {Stockdale}, C.~J.,
  {Kelley}, M.~T., {Sramek}, R.~A., {Van Dyk}, S.~D., \& {Marcaide}, J.~M.
  2007, \apj, 671, 1959

\end{thebibliography}

\clearpage

\end{document}